\documentclass[12pt, notitlepage,preprintnumbers,superscriptaddress,nofootinbib,aps,prd,floatfix]{revtex4-2}
\usepackage[a4paper,top=1.5cm,bottom=2cm,left=1.5cm,right=1.5cm,bindingoffset=0mm]{geometry}
\pdfoutput=1
\usepackage{float}
\usepackage{physics,slashed}
\usepackage{amsfonts,amsmath,amssymb}
\usepackage{gensymb}%degree
\usepackage{mathrsfs}
\usepackage{bm}% bold math

\usepackage{epsfig}
\usepackage{graphicx}   % need for figures
\usepackage{subfigure}  % use for side-by-side figures
\usepackage{verbatim}   % useful for program listings
\usepackage{xcolor}      % use if color is used in text
\usepackage[linktocpage]{hyperref}
\usepackage{url}
\definecolor{nicered}{rgb}{0.5,0.,0.}
\definecolor{nicegreen}{rgb}{0.,0.5,0.}
\definecolor{niceblue}{rgb}{0.,0.,0.5}
\hypersetup{colorlinks,citecolor=nicegreen,linkcolor=nicered,urlcolor=niceblue}
\usepackage{array}
\usepackage{dcolumn}% Align table columns on decimal point
\usepackage{multirow}
\usepackage{cprotect}
\usepackage{framed}
\usepackage{setspace}
\usepackage{enumitem,kantlipsum}
\usepackage{ulem}
\setlist{nosep}

\usepackage{tikz-feynman}
\tikzfeynmanset{compat=1.1.0}

\usepackage{tikzsymbols}

\newcommand\barparenb[1]{\overset{%
   \scalebox{0.4}{$(\mkern-1mu-\mkern-1mu)$}}{#1}}
\newcommand{\GeV}{\textrm{GeV}}
\newcommand{\TeV}{\textrm{TeV}}

\newcommand{\zp}{$Z^\prime~$}

\begin{document}
\preprint{PITT-PACC-2317, IP/BBSR/2023-09}
\title{Searching for Heavy Leptophilic $Z'$: \\
from Lepton Colliders to Gravitational Waves}

\author{Arnab Dasgupta}
\email{arnabdasgupta@pitt.edu}
\affiliation{Pittsburgh Particle Physics, Astrophysics, and Cosmology Center, Department of Physics and Astronomy, University of Pittsburgh, Pittsburgh, PA 15206, USA\looseness=-1}
\author{P. S. Bhupal Dev}
\email{bdev@wustl.edu}
\affiliation{Department of Physics and McDonnell Center for the Space Sciences, Washington University, St. Louis, MO 63130, USA\looseness=-1}
\author{Tao Han}
\email{than@pitt.edu}
\affiliation{Pittsburgh Particle Physics, Astrophysics, and Cosmology Center, Department of Physics and Astronomy, University of Pittsburgh, Pittsburgh, PA 15206, USA\looseness=-1}
\author{Rojalin Padhan}
\email{rojalin.p@iopb.res.in}
\affiliation{Pittsburgh Particle Physics, Astrophysics, and Cosmology Center, Department of Physics and Astronomy, University of Pittsburgh, Pittsburgh, PA 15206, USA\looseness=-1}
\affiliation{Institute of Physics, Sachivalaya Marg, Bhubaneswar 751005, Odisha, India\looseness=-1}
\affiliation{Homi Bhabha National Institute, BARC Training School Complex, Anushakti Nagar, Mumbai 400094, India\looseness=-1}
\author{Si Wang}
\email{siw34@pitt.edu}
\affiliation{Pittsburgh Particle Physics, Astrophysics, and Cosmology Center, Department of Physics and Astronomy, University of Pittsburgh, Pittsburgh, PA 15206, USA\looseness=-1}
\author{Keping Xie}
\email{xiekeping@pitt.edu}
\affiliation{Pittsburgh Particle Physics, Astrophysics, and Cosmology Center, Department of Physics and Astronomy, University of Pittsburgh, Pittsburgh, PA 15206, USA\looseness=-1}
\affiliation{Department of Physics and Astronomy, Michigan State University, East Lansing, MI 48824, USA}
%%\keywords{Specific BSM Phenomenology, New Gauge Interactions,Cosmology of Theories BSM,Phase Transitions in the Early Universe,Early Universe Particle Physics}
%%%%%%%%%%%%%%%%%%%%%%%%%%%%%%%%%%
\begin{abstract}
We study the phenomenology of leptophilic $Z'$ gauge bosons at the future high-energy $e^+e^-$ and $\mu^+\mu^-$ colliders, as well as at the gravitational wave observatories. The leptophilic $Z'$ model, although well-motivated, remains largely unconstrained from current low-energy and collider searches for $Z'$ masses above ${\cal O}(100~{\rm GeV})$, thus providing a unique opportunity for future lepton colliders. Taking $U(1)_{L_\alpha-L_\beta}~(\alpha,\beta=e,\mu,\tau)$ models as concrete examples, we show that future $e^+e^-$ and $\mu^+\mu^-$ colliders with multi-TeV center-of-mass energies provide unprecedented sensitivity to heavy leptophilic $Z'$ bosons. Moreover, if these $U(1)$ models are classically scale-invariant, the phase transition at the $U(1)$ symmetry-breaking scale tends to be strongly first-order with ultra-supercooling, and leads to observable stochastic gravitational wave signatures. We find that the future sensitivity of gravitational wave observatories,  such as advanced LIGO-VIRGO and Cosmic Explorer, can be complementary to the collider experiments, probing higher $Z'$ masses up to ${\cal O}(10^4~{\rm TeV})$, while being consistent with naturalness and perturbativity considerations. 
\end{abstract}
%%%%%%%%%%%%%%%%%%%%%%%%%%%%%%%%%%
\maketitle
\tableofcontents
%%%%%%%%%%%%%%%%%%%%%%%%%%%%%%%%%%
\section{Introduction}
\label{sec:intro}
%%%%%%%%%%%%%%%%%%%%%%%%%%%%%%%%%%
High-energy colliders have immensely enriched our understanding of nature at the most fundamental level. The Large Hadron Collider (LHC), in particular, has enabled the exploration of energy scales as high as several TeV, and has consolidated the robustness of the Standard Model (SM). However, the primary pursuit of the LHC, namely, finding beyond-the-SM (BSM) phenomena, remains elusive. 
Despite tremendous efforts of theoretical and experimental works to address the empirical and theoretical shortcomings of the SM, no sign of BSM physics has been observed so far, and stringent bounds (up to several TeV) have been placed on their mass scale~\cite{ParticleDataGroup:2022pth}.  
Given all these constraints, one might wonder whether the scale of BSM physics must lie {\it beyond} the energy scales accessible at the LHC. 

There is one notable loophole in this general argument for hadron colliders, 
{\it viz.},~if the new resonance is electrically neutral and couples only to the SM leptons at leading order. Most of the LHC (and Tevatron) bounds coming from resonance searches do not directly apply to such a neutral {\it leptophilic} sector. The relevant collider constraints in this case mainly come from LEP (or from LHC via higher-order processes), and are generally much weaker than the direct LHC constraints applicable for hadrophilic resonances. The LEP constraints from resonance searches are typically around the 100 GeV scale, limited by its center-of-mass energy of $\sqrt s=209$ GeV, whereas the LEP contact interaction bounds for heavier particles are at most in the TeV  range (for ${\cal O}(1)$ couplings)~\cite{Electroweak:2003ram, ALEPH:2013dgf}. Therefore, future lepton colliders, such as the proposed $e^+e^-$ colliders like ILC~\cite{ILC:2013jhg}, CEPC~\cite{CEPCStudyGroup:2018ghi}, FCC-ee~\cite{FCC:2018evy}, and CLIC~\cite{CLICdp:2018cto}, or a high-energy $\mu^+\mu^-$ collider~\cite{Delahaye:2019omf,Long:2020wfp, MuonCollider:2022xlm, Accettura:2023ked},  are uniquely capable of probing leptophilic BSM particles to unprecedented mass and coupling values. In this paper, we will focus on leptophilic neutral gauge bosons ($Z'$) as a case study to illustrate this point.   

At first sight, new neutral gauge bosons, especially coupling only to leptons and not to quarks at tree level, may look artificial. However, there are good symmetry reasons that can motivate such a scenario. First of all, the presence of additional Abelian symmetries like $U(1)$ is quite natural and can be motivated by Grand Unified Theories, string compactifications, extra dimensional models, solutions of the gauge hierarchy problem, and so on~\cite{Langacker:2008yv}, which always come with the corresponding $Z'$ bosons after the $U(1)$-symmetry breaking. As for the leptophilic nature of $Z'$, it is well-known that the classical SM Lagrangian already contains the accidental global symmetry $U(1)_{e}\times U(1)_{\mu}\times U(1)_{\tau}$, associated with conserved lepton number for each family. A simple $U(1)$ gauge extension of the SM allows us to promote $U(1)_{L_\alpha-L_\beta}$ (where $\alpha,\beta=e,\mu,\tau$ with $\alpha\neq \beta$) to an anomaly-free local gauge symmetry~\cite{Foot:1990mn, He:1991qd, Foot:1994vd}. The associated $Z'$ gauge bosons are {\it naturally} leptophilic, with couplings to quarks induced only at the loop level. Therefore, the most stringent dijet constraints on heavy $Z'$ coming from Tevatron and LHC~\cite{Dobrescu:2021vak} are not applicable in this scenario, thus opening up a large swath of parameter space in the $Z'$ mass-coupling plane.\footnote{Such a leptophilic $Z'$ can also serve as a portal to the dark sector, with very interesting phenomenology~\cite{Fox:2008kb, Kopp:2009et, Agrawal:2014ufa, Bell:2014tta, Alves:2015pea,Blanco:2019hah}.} In this paper, we target this currently unexplored leptophilic $Z'$ parameter space and study the future lepton collider prospects of probing this well-motivated BSM scenario. We consider all possible production channels (both resonance and off-resonance, associated production with photons, and final-state radiation) to carve out the future lepton collider sensitivity in the $Z'$ parameter space. Taking $\sqrt s=3$ TeV electron/muon collider with an integrated luminosity of 1 ab$^{-1}$ as a case study, we find up to three orders of magnitude improvement over the existing constraints for the $Z'$ coupling sensitivity reach over a broad mass range.

Another interesting and complementary aspect of the leptophilic $U(1)$ models we study here is the cosmological phase transition of the $U(1)$-symmetry-breaking scalar field. If the symmetry is classically conformal~\cite{Meissner:2006zh}, the tree-level potential is flat due to scale-invariance, and thermal corrections can easily dominate and make the phase transition strongly first order~\cite{Farzinnia:2014yqa}, leading to a potentially observable stochastic  gravitational wave (GW) signal. The conformal invariance is motivated as a possible solution to the gauge hierarchy problem in the SM. It is well-known that the fermion masses in the SM are protected by chiral symmetry, i.e. in the limit of $m_f\to 0$, radiative corrections $\delta m_f$ also vanish to all orders in perturbation theory~\cite{tHooft:1979rat}. However, this is not the case for the Higgs boson mass, where the radiative correction $\delta m_h$ does not vanish in the limit of $m_h\to 0$. It leads to the puzzle of the stabilization of the weak scale against large radiative corrections -- the so-called gauge hierarchy problem~\cite{Gildener:1976ai, Weinberg:1978ym}. This can, in principle, be evaded in a classical conformal theory~\cite{Kawamura:2013kua}, where all the mass scales (including the electroweak scale) are generated by dimensional transmutation using the Coleman-Weinberg mechanism~\cite{Coleman:1973jx}.  However, this mechanism cannot be applied directly to the SM Higgs sector since the predicted Higgs mass is too low, and moreover, the Coleman-Weinberg effective potential in the SM becomes unbounded from below~\cite{Fujikawa:1978ru}. Instead, the $U(1)$ models provide a viable alternative to realize the conformal invariance~\cite{Iso:2009ss, Iso:2009nw, Iso:2012jn, Oda:2015gna, Guo:2015lxa,
Das:2016zue}. We study the interconnection and the complementarity of our collider signals with the GW signal in classically conformal versions of the leptophilic $U(1)_{L_\alpha-L_\beta}$ models. We show that the current GW data from aLIGO-aVIRGO~\cite{KAGRA:2021kbb} already excludes a portion of the $U(1)_{L_\alpha-L_\beta}$ model parameter space at high $M_{Z'}$ values not accessible to colliders, whereas the next-generation GW experiments in the mHz-kHz regime, such as $\mu$ARES~\cite{Sesana:2019vho}, LISA~\cite{LISA:2017pwj}, DECIGO~\cite{Seto:2001qf}, BBO~\cite{Corbin:2005ny}, ET~\cite{Punturo:2010zz}, and CE~\cite{Reitze:2019iox} will further extend the sensitivity reach to our $Z'$ parameter space of interest.   We also comment on the possibility of explaining the recent Pulsar Timing Array observations of stochastic GW at nHz frequencies~\cite{NANOGrav:2023gor, Antoniadis:2023ott, Reardon:2023gzh, Xu:2023wog} in these models. 

The rest of the paper is organized as follows. In Section~\ref{sec:collider}, we briefly discuss the leptophilic $Z'$ models under consideration and summarize the existing bounds on the model parameter space. In Section~\ref{sec:pheno}, we analyze various production channels for the $Z'$ boson at future lepton colliders, and summarize the sensitivity limits. In Section~\ref{sec:GW}, we study the GW signals in a conformal version of the $U(1)$ models. Our conclusions are given in Section~\ref{sec:con}.  

%%%%%%%%%%%%%%%%%%%%%%%%%%%%%%%%%%%%%% 
\section{Leptophilic $Z'$ and current constraints}
\label{sec:collider}
%%%%%%%%%%%%%%%%%%%%%%%%%%%%%%%%%%
\subsection{$U(1)_{L_\alpha-L_\beta}$ models}
%%%%%%%%%%%%%%%%%%%%%%%%%%%%%%%%%%
Within the particle content of the SM, it is possible to gauge one of the three combinations of $L_\alpha-L_\beta$ ($\alpha,\beta=e,\mu,\tau$), without introducing an anomaly~\cite{Foot:1990mn, He:1991qd, Foot:1994vd}. This surprising feature is the main motivation behind the minimal BSM framework considered here, with the SM gauge symmetry extended by an extra leptophilic $U(1)_{L_\alpha-L_\beta}$, such that only two lepton flavors $\alpha$ and $\beta$ are oppositely charged, while all other SM fields are neutral under this $U(1)$ gauge symmetry. It is interesting to note that although we can formally consider the anomaly-free combination $U(1)_{L_e-L_\mu}\times U(1)_{L_\mu-L_\tau}\times U(1)_{L_e-L_\tau}$, the decomposition $L_e-L_\tau=(L_e-L_\mu)+(L_\mu-L_\tau)$ shows that not all of their generators are independent and only two of the lepton number differences can be gauged. The question then arises of which subgroup should be chosen. The $L_\mu-L_\tau$ option is the most popular of the three, because (a) it predicts the neutrino mass matrix to be $L_\mu-L_\tau$ symmetric, which (with a small symmetry-breaking effect) fits the observed neutrino oscillation data very well~\cite{Binetruy:1996cs,Bell:2000vh, Lam:2001fb, Choubey:2004hn,Asai:2017ryy}, and (b) it provides a simple solution to the muon $g-2$ anomaly~\cite{Gninenko:2001hx,Baek:2001kca,Ma:2001md}. See Refs.~\cite{Ota:2006xr,Heeck:2011wj, Foldenauer:2018zrz,Escudero:2019gzq,Joshipura:2019qxz,Dev:2020drf, Zhang:2020fiu,Huang:2021nkl,Araki:2021xdk,Amaral:2021rzw,Buras:2021btx, Borah:2021mri,Drees:2021rsg,Hapitas:2021ilr,Asai:2021wzx, Cheng:2021okr, Heeck:2022znj, Eijima:2023yiw, Chen:2023mep} for various phenomenological studies of the $U(1)_{L_\mu-L_\tau}$ model. The other two combinations $U(1)_{L_e-L_\mu}$ and $U(1)_{L_e-L_\tau}$ have also been considered in various contexts~\cite{Araki:2012ip,Heeck:2018nzc,Ballett:2019xoj,Asai:2019ciz,Dev:2021xzd,Ardu:2022zom}. Here we will be agnostic of the possible flavor-gauge connection, and will primarily focus on the future lepton collider phenomenology of each of the three combinations of $U(1)_{L_\alpha-L_\beta}$ separately. 

In the minimal setup, the $U(1)_{L_\alpha-L_\beta}$ gauge symmetry is spontaneously broken by the vacuum expectation value (VEV) of a complex scalar field $\Phi$, which is neutral under the SM gauge group but charged under  $U(1)_{L_\alpha-L_\beta}$; see Table~\ref{tab:Model} for the charge assignments in the $L_\mu-L_\tau$ case as an example.   The relevant terms in the effective  Lagrangian are given by 
\begin{equation}
-\mathcal{L}_{L_\alpha-L_\beta} \supset g' Z'_\mu (
\bar{L}_{\alpha}\gamma^\mu L_\alpha-\bar{L}_\beta\gamma^\mu L_\beta+\bar{\ell}_{R\alpha}\gamma^\mu \ell_{R\alpha}
-\bar{\ell}_{R\beta}\gamma^\mu e_{\ell\beta})
+\frac{1}{2}M_{Z'}^2Z'_\mu Z'^{\mu},
\end{equation} 
where the $Z'$ mass is given by $M_{Z'}=2g'v_\Phi$ (with $v_\Phi$ being the VEV of $\Phi$; see Section~\ref{sec:potential}).  

\begin{table}
\centering
\begin{tabular}{|c||c|c|c|c|c|c|c|c|}
\hline
 Gauge group & $L_{e}$& $L_{\mu}$ & $L_{\tau}$ & $e_R$ & $\mu_R$ & $\tau_R$& $H$ & $\Phi$\\
\hline\hline
$SU(3)_c$ & {\bf 1} &{\bf 1} & {\bf 1}& {\bf 1} &{\bf 1} & {\bf 1}& {\bf 1} &{\bf 1} \\
$SU(2)_L$ &  {\bf 2} &{\bf 2} & {\bf 2}& {\bf 1} &{\bf 1} & {\bf 1}& {\bf 2} &{\bf 1}\\
$U(1)_Y$ & $-\frac{1}{2}$ &$-\frac{1}{2}$ & $-\frac{1}{2}$& $-1$ & $-1$ & $-1$ & $\frac{1}{2}$ &0\\
$U(1)_{L_\mu-L_\tau}$ & 0 &1 & $-1$ & 0 &1 & $-1$ & 0 &2\\
\hline
\end{tabular}
\caption{Particle content and charges in the $U(1)_{L_\mu-L_\tau}$ model as an example of the $U(1)_{L_\alpha-L_\beta}$ models. Here $L_\alpha=(\nu_L,\ell_L)^T_\alpha$ and $H$ stand for the $SU(2)_L$ lepton and Higgs doublets, respectively.}
\label{tab:Model}
\end{table}

Here we do not consider the kinetic mixing of  $Z'_\mu$ with the SM $B_\mu$ field, or mass mixing with the $Z_\mu$ field, i.e., the terms in the Lagrangian~\cite{Galison:1983pa, Holdom:1985ag, Babu:1997st}
\begin{equation}
    {\cal L}_{\rm mix} = -\epsilon Z'^{\mu\nu}B_{\mu\nu}+\delta M^2 Z'^{\mu}Z_\mu \, ,
\end{equation}
where $B_{\mu\nu}$ and $Z'_{\mu\nu}$ are the field-strength tensors for $U(1)_Y$ and $U(1)_{L_\alpha-L_\beta}$, respectively. The mass mixing term is naturally absent in our case because the Higgs boson of one group is not charged under the second group; see Table~\ref{tab:Model}. The kinetic mixing term can be forbidden at the tree level by the introduction of a discrete symmetry $\alpha\leftrightarrow\beta$~\cite{Foot:1994vd, Ibe:2016dir} under which $B_{\mu\nu}\to B_{\mu\nu}$ and $Z'_{\mu\nu}\to -Z'_{\mu\nu}$. But this is held only by the gauge interaction part of the QED and is softly broken by the lepton mass terms ($m_\alpha\neq m_\beta$), which will generate an unavoidable finite kinetic mixing by radiative corrections. It can be evaluated from the mixing between $B_\mu$ and $Z'_\mu$ induced by lepton loop as~\cite{Araki:2017wyg}:
\begin{eqnarray}
&& \includegraphics[width=5cm]{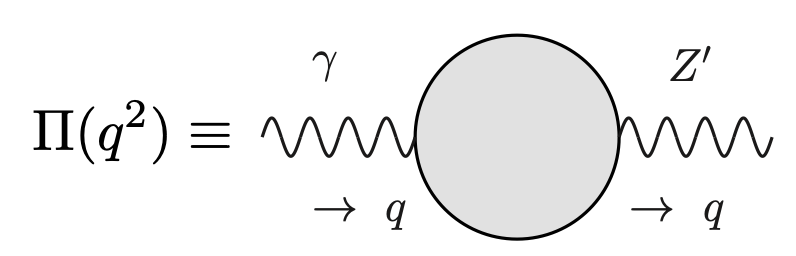} \nonumber \\
&& \qquad \quad = \frac{8eg'}{16\pi^2}\int_0^1 dx~x(1-x)\log\left[\frac{m_{\ell_\beta}^2-x(1-x)q^2}{m_{\ell_\alpha}^2-x(1-x)q^2} \right] \, .
\end{eqnarray}
Note that this contribution is finite. For our subsequent collider study, the relevant parameter space is $M_{Z'}\gtrsim 100$ GeV. Restricting the $q^2$ scale accordingly, we find the induced kinetic mixing to be $\epsilon\lesssim 4\times 10^{-4}$ for $U(1)_{L_\alpha-L_\tau}$ and $\lesssim 2\times 10^{-6}$ for  $U(1)_{L_e-L_\mu}$. Therefore, we can safely neglect it in our analysis. 

Before we move on to the experimental constraints and lepton collider searches for the new $U(1)$ gauge boson, we first present its decay and lifetime properties.
The partial decay width of $Z'\to\ell^+\ell^-$ for a single lepton flavor is given by
\begin{equation}\label{eq:PW}
\Gamma(Z'\to\ell^+\ell^-)=\frac{g'^2}{12\pi}M_{Z'}\left(1+\frac{2m_\ell^2}{M_{Z'}^2}\right)
\sqrt{1-\frac{4m_\ell^2}{M_{Z'}^2}} \, .
\end{equation}
The $U(1)_{L_\alpha-L_\beta}$ gauge involves only two lepton flavors, with the allowed $Z'$ decay channels as\footnote{The coupling of $Z'$ to the third lepton flavor is induced at loop level via $\gamma-Z'$ mixing, which is however suppressed by $m^2_{\ell_{\alpha,\beta}}/M^2_{Z'}$~\cite{Araki:2017wyg}.}
\begin{equation}
Z'\to \ell^+_\alpha\ell^-_{\alpha},\ \ell^+_\beta\ell^-_{\beta},\ 
\nu_{\alpha}\bar{\nu}_{\alpha},\ \nu_{\beta}\bar{\nu}_{\beta} \, .
\end{equation}
Considering 
$M_{Z'}\gg m_\ell$, we can safely ignore the lepton mass in Eq.~\eqref{eq:PW}, and obtain the total width of $Z'$ as 
\begin{equation}
\label{eq:decay}
\Gamma_{Z'}\simeq \frac{(2N_\ell+N_\nu)g'^2}{24\pi}M_{Z'} = \frac{g'^2}{4\pi}M_{Z'}\, ,
\end{equation}
where $N_{\ell}=N_{\nu}=2$ for two lepton flavors. Note that each neutrino flavor only contributes half to Eq.~(\ref{eq:PW}) because of its left-handed chirality.
Here we do not consider the potential $Z'$ interactions with right-handed neutrinos or with dark sector.

\begin{figure}[t!]
    \centering    \includegraphics[width=0.49\textwidth]{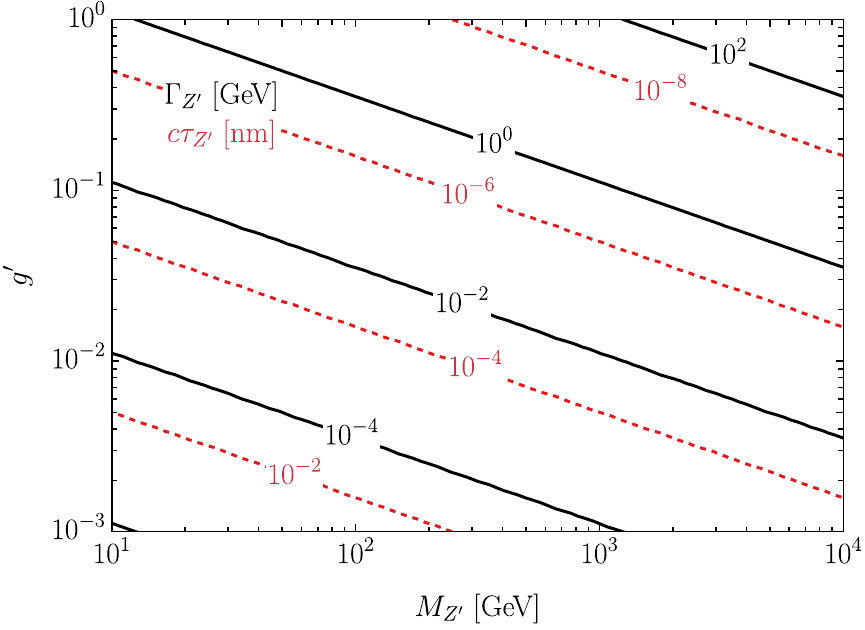}
    \caption{Dependence of $Z'$ decay width $\Gamma_{Z'}$ (solid contours) and the corresponding proper decay length $c\tau_{Z'}$ (dashed contours) on its mass $M_{Z'}$ and gauge coupling $g'$ in the $U(1)_{L_\alpha-L_\beta}$ model. 
    }
    \label{fig:decay}
\end{figure}

In Fig.~\ref{fig:decay}, we present the $Z'$ decay width with respect to its mass $M_{Z'}$ and the corresponding gauge coupling $g'$. We see that for the parameter region of our current interest, {\it i.e.}, $M_{Z'}\in[10,10^{4}]~\GeV$ with $g'\in[10^{-3},1]$, the $Z'$ decay width spans the range $\sim[10^{-6},10^{2}]~\GeV$. 
The proper decay length can be estimated by
\begin{equation}
c\tau_{Z'} =\frac{c}{\Gamma_{Z'}}
= 2.5 \times 10^{-4}\ \textrm{nm}\left({10^{-2} \over g'}\right)^2\ 
\left({ {\rm 100\ GeV} \over M_{Z'}}\right)  \, ,
\end{equation}
shown as red dashed lines in Fig.~\ref{fig:decay}.
We see that $Z'$ will decay promptly in the parameter space of our interest, which can potentially leave direct or indirect signals at colliders. Based on this knowledge, we will focus on the $Z'$ production followed by its prompt decay in the rest of this work.
%%%%%%%%%%%%%%%%%%%%%%%%
%%%%%%%%%%%%%%%%%%%%%%%%%%%%%%%%%%
\subsection{Current laboratory bounds on the model parameters}
\label{sec:bounds}
\begin{figure}[t!]
    \centering
    \includegraphics[width=0.49\textwidth]{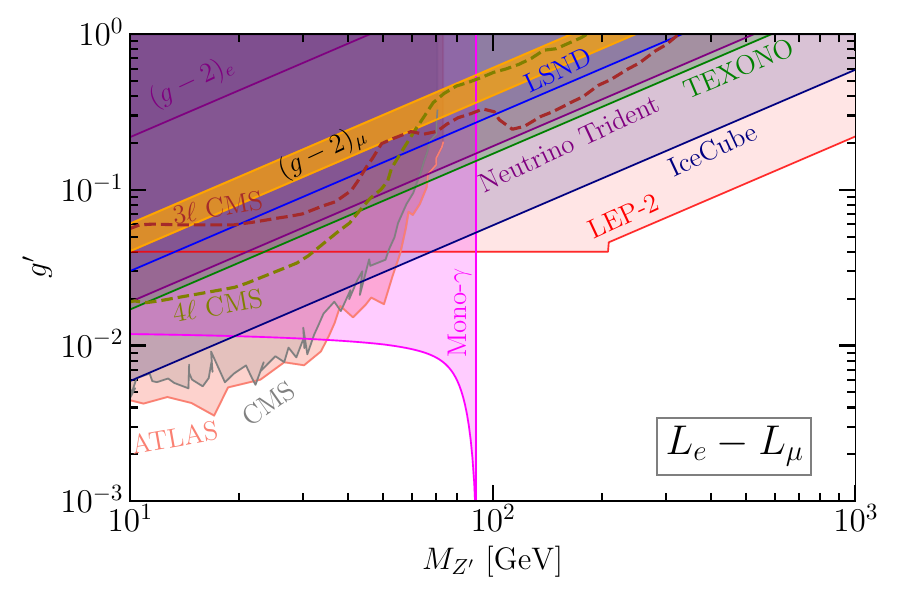}     
    \includegraphics[width=0.49\textwidth]{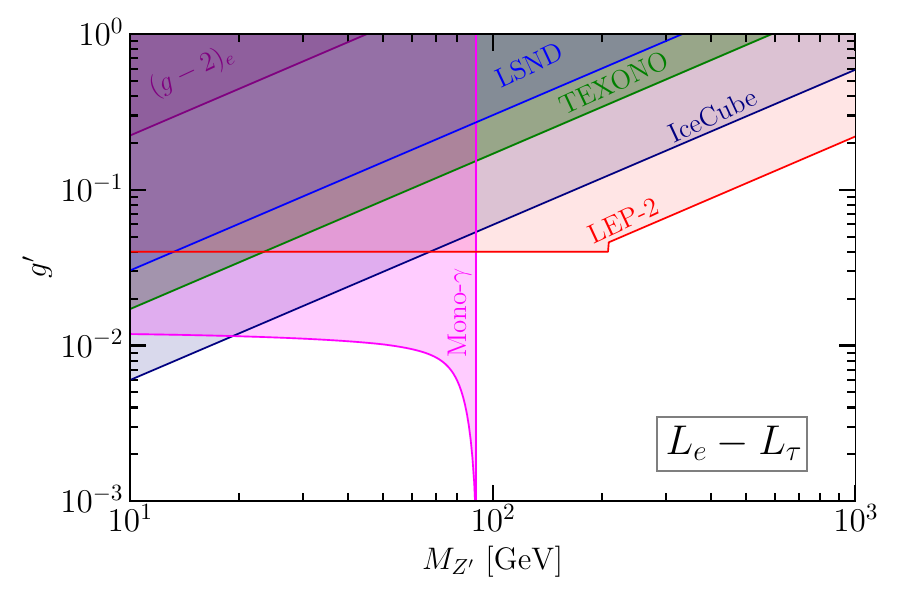}
    \includegraphics[width=0.49\textwidth]{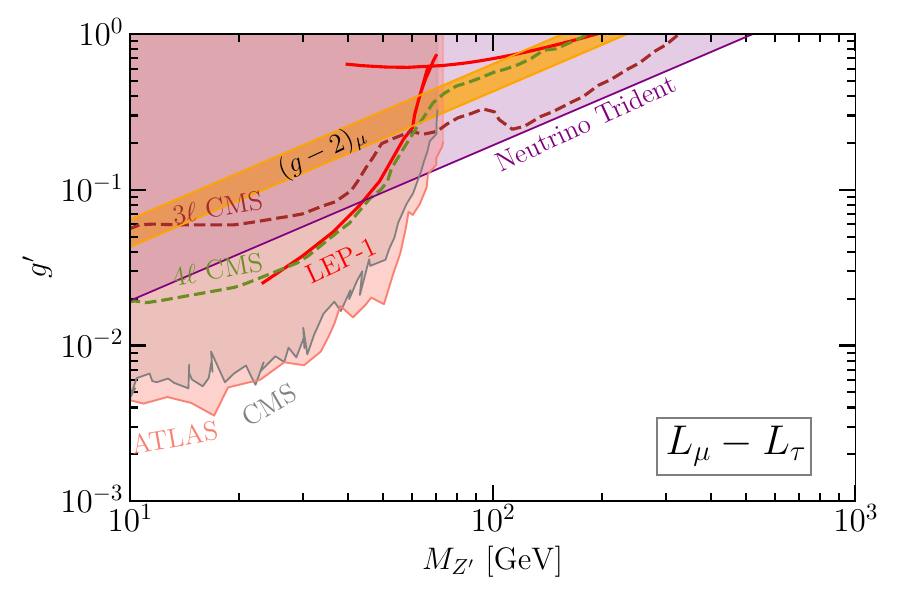}    
    \caption{Existing $95\%$ CL exclusion bounds (except IceCube, which is at $90\%$ CL) on the gauge boson mass $M_{Z'}$ and coupling $g'$ in the $U(1)_{L_\alpha-L_\beta}$ models. See Section~\ref{sec:bounds} for details.
}
\label{fig:exist}
\end{figure}

%%%%%%%%%%%%%%%%%%%%%%%%%%%%%%%%%%
In this section, we summarize the existing constraints on the $U(1)_{L_\alpha-L_\beta}$ gauge boson mass~$M_{Z^\prime}$ and coupling~$g^\prime$, as shown in Fig.~\ref{fig:exist} and explained below. This is the most comprehensive set of constraints on these models available to date for our region of interest, {\it i.e.}, for $M_{Z'}>10$ GeV; for    other constraints relevant in the low-mass regime, see {\it e.g.}, the summary plots in Refs.~\cite{Ilten:2018crw, Bauer:2018onh}. All the constraints shown here are at $95\%$ confidence level (CL), unless otherwise specified.

\begin{itemize}[listparindent=1.5em]
\setlength\itemsep{5pt} 
\item 
$\bm{(g-2)_\ell}$: The new interaction between $Z'$ and the SM leptons will give a contribution to the anomalous magnetic dipole moment $a_\ell=(g-2)_\ell/2$ of the corresponding leptons as~\cite{Leveille:1977rc, Fayet:2007ua} 
\begin{equation}
   \Delta a_\ell\simeq \frac{g'^2}{4\pi^2}\int_0^1 \dd x \frac{m_\ell^2 x^2(1-x)}{m_\ell^2x^2+M_{Z'}^2(1-x)}\simeq \frac{g^{\prime2} m_\ell^2}{12\pi^2 M_Z^{\prime 2}} \,  
   \label{eq:gm2}
\end{equation}
for $M_{Z'}\gg m_\ell$. For $a_e$, although the experimental value is known very precisely~\cite{Fan:2022eto}, the SM prediction~\cite{Aoyama:2019ryr} relies on the measurement of the fine-structure constant using the recoil velocity/frequency of atoms that absorb a photon, and currently there is a $5.5\sigma$ discrepancy between the measurements using Rubidium-87~\cite{Morel:2020dww} and Cesium-133~\cite{Parker:2018vye}. When translated into $\Delta a_e$, they give   
%Currently, the latest measurement of the electron anomalous magnetic moment $a_e$ was based on the recoil velocity/frequency of  atoms that absorb a photon. These experiments give deviations from the Standard Model prediction as
\begin{equation}
\Delta a_{e}=\left\{\begin{array}{ll}(4.8\pm3.0 )\times 10^{-13} & ({\rm{Rb}})\\
(-8.8\pm3.6 )\times 10^{-13} &({\rm{Cs}})  \end{array} \right. .
\end{equation}
Since the correction from $Z'$ loop is of positive sign (cf.~Eq.~\eqref{eq:gm2}), we use the Rb-measurement and show the $95\%$ CL upper limit in Fig.~\ref{fig:exist}  (labeled as $(g-2)_e$), which goes like $g'\lesssim 2.2\times 10^{-2}M_{Z'}/{\rm GeV}$, and is applicable to the $U(1)_{L_e-L_\alpha}$ models. 

As for $a_\mu$, the Muon $g-2$ Collaboration combined the recent Fermilab measurements~\cite{Muong-2:2021ojo, Muong-2:2023cdq} with the old Brookhaven E821 result~\cite{Muong-2:2006rrc}, and obtained a $5.0\sigma$ deviation from the world average of the SM expectation~\cite{Aoyama:2020ynm}:
\begin{equation} 
\Delta a_{\mu}=(2.49\pm0.48)\times 10^{-9}\, . 
\label{eq:gmmu}
\end{equation}
This discrepancy is reduced to only about $2\sigma$, if the ab initio lattice simulation result from BMW Collaboration~\cite{Borsanyi:2020mff} is used for the SM result. However, this claim is being independently verified by other lattice groups~\cite{Ce:2022kxy, ExtendedTwistedMass:2022jpw,FermilabLatticeHPQCD:2023jof,Blum:2023qou,Wittig:2023pcl}, and until the issue is settled, we prefer to use Eq.~\eqref{eq:gmmu} to derive the $95\%$ CL-preferred region in the $Z'$ parameter space, as shown by the orange-shaded band in Fig.~\ref{fig:exist}. Moreover, it gives a $5\sigma$ exclusion bound of $g'\lesssim 7.6\times 10^{-3}M_{Z'}/{\rm GeV}$, applicable to both $U(1)_{L_e-L_\mu}$ and $U(1)_{L_\mu-L_\tau}$ models. 

In comparison, $a_\tau$ is known relatively poorly. The current experimental limits are 
\begin{equation}
    a_\tau = \left\{\begin{tabular}{ll}
$[-0.052,0.013]$ & ( DELPHI 95\% CL  \cite{DELPHI:2003nah} )  \\ 
$[-0.057,0.024]$ & ( ATLAS 95\%  CL \cite{ATLAS:2022ryk} ) \\  
$0.001^{+0.055}_{-0.089} $ & ( CMS 68\% CL  { \cite{CMS:2022arf} })
\end{tabular}\right. .
\end{equation} 
A global analysis of LEP and SLD data on the $\tau$-lepton pair production in an effective field theory framework yields a slightly tighter constraint: $a_\tau=[-0.007,0.005]$ at $95\%$ CL level~\cite{Gonzalez-Sprinberg:2000lzf}. An even better bound of $|a_\tau|<1.8\times 10^{-3}$ at $95\%$ CL was obtained recently using the full LHC Run 2 data on tau-pair production~\cite{Haisch:2023upo}. Taking the leading one-loop QED result for the SM prediction, $a_\tau^{\rm SM}=\alpha/2\pi\simeq 1.16\times 10^{-3}$~\cite{Schwinger:1948iu}, we get a $95\%$ CL bound on $|\Delta a_\tau|<6.4\times 10^{-4}$, which translates into a rather weak bound of $g'\lesssim 0.15M_{Z'}/{\rm GeV}$, applicable to $U(1)_{L_\alpha-L_\tau}$ models. This, however, falls outside the range shown in Fig.~\ref{fig:exist}.

\item\textbf{Neutrino trident production:} Another strong bound comes from the production of a muon-antimuon pair in the scattering of muon neutrinos in the Coulomb field of a target nucleus, {\it e.g.}, neutrino trident production~\cite{Altmannshofer:2014pba}. A combination of measurements of the trident cross-section from CHARM-II~\cite{CHARM-II:1990dvf}, CCFR~\cite{CCFR:1991lpl} and NuTeV~\cite{NuTeV:1998khj} imposes a bound of $g'\lesssim 1.9\times 10^{-3}M_{Z'}/{\rm GeV}$ on the $U(1)_{L_e-L_\mu}$ and $U(1)_{L_\mu-L_\tau}$ models~\cite{Altmannshofer:2016jzy}, as shown in Fig.~\ref{fig:exist} by the purple shaded region. This trident bound rules out the region preferred by the $(g-2)_\mu$ anomaly in the entire high mass range.\footnote{There still exists some allowed parameter space in the $U(1)_{L_\mu-L_\tau}$ model that can explain the  $(g-2)_\mu$ anomaly for a lighter $M_{Z'}\sim 10$--100 MeV and $g'\sim 10^{-3}$~\cite{Amaral:2021rzw}. This region can be probed in low-energy experiments like NA64-e~\cite{NA64:2022rme}.} 

\item \textbf{Neutrino scattering:} The interaction between \zp and electrons contributes to the elastic $\barparenb{\nu}_e+e^-\to \barparenb{\nu}_e+e^-$ scattering process in $U(1)_{L_e-L_\alpha}$ models~\cite{Bell:2014tta, Dev:2021xzd}. Using the $\nu_e-e$ elastic scattering cross-section measurement from LSND~\cite{LSND:2001akn}, we obtain a $95\%$ CL bound of $g'\lesssim 3\times 10^{-3}M_{Z'}/{\rm GeV}$~\cite{Bell:2014tta}, as shown in Fig.~\ref{fig:exist} by blue shaded area.
Similarly, the $\bar{\nu}_ee^-$ scattering cross-section has been measured by TEXONO~\cite{TEXONO:2009knm}, which sets a limit of $g'\lesssim 1.7\times 10^{-3}M_{Z'}/{\rm GeV}$~\cite{Bilmis:2015lja,Bauer:2018onh,Dev:2021xzd}. This constraint is shown by the green shaded region in Fig.~\ref{fig:exist}.
%\BD{This bound looks slightly different for $L_e-L_\mu$ versus $L_e-L_\tau$. Please check}\BD{How did you do it in the current version? And what is the reason for the difference?}\sw{solved}

\item \textbf{IceCube:} The $Z'$ couplings with SM neutrinos and electrons will introduce non-standard neutrino interactions (NSI), which impact the neutrino-matter effective potential~\cite{Wise:2018rnb, Proceedings:2019qno}. The NSI can be parametrized as $\varepsilon_{\alpha\beta}=g'^2/\sqrt 2 G_F M^2_{Z'}$, where $G_F\simeq 1.166\times 10^{-5}~{\rm GeV}^{-2}$ is the Fermi constant. Using the 8-year IceCube DeepCore data, a preliminary $90\%$ CL bound on $|\varepsilon_{\tau\tau}-\varepsilon_{\mu\mu}|\leq 2.1\times 10^{-2}$ has been obtained~\cite{IceCube:2022pbe}, which is stronger by a factor of few from the published limits from IceCube~\cite{IceCubeCollaboration:2021euf} and ANTARES~\cite{ANTARES:2021crm}, and by almost an order of magnitude stronger than the Super-Kamiokande limit~\cite{Super-Kamiokande:2011dam}. The NSI constraint translates into a bound of $g'\lesssim 5.9\times 10^{-4}M_{Z'}/{\rm GeV}$, which applies to the $U(1)_{L_e-L_\mu}$ and $U(1)_{L_e-L_\tau}$ models, as shown in Fig.~\ref{fig:exist} by the navy blue shaded region. 
\item \textbf{LEP:} The coupling of \zp to electrons is strongly constrained by the measured cross-section for the processes $e^+e^-\to \ell^+\ell^-$ at the LEP-2 experiment~\cite{Electroweak:2003ram}. For $M_{Z^\prime}>\sqrt s= 209~\GeV$, we use the four-fermion contact interaction bounds, expressed as $g^\prime\le  0.044 M_Z^\prime/(200~\GeV)$~\cite{Buckley:2011vc}. For $M_{Z^\prime}<209~\GeV$, the four-fermion description is no longer valid, and a conservative limit of $g^\prime\le  0.04$ is used~\cite{Bell:2014tta}. These limits apply to the $U(1)_{L_e-L_\mu}$ and $U(1)_{L_e-L_\tau}$ models, as shown in Fig.~\ref{fig:exist} by the red shaded regions. For the $U(1)_{L_\mu-L_\tau}$ model, the LEP-1 measurement of the four-fermion final-state at the $Z$ pole~\cite{ALEPH:1994bie} was used to derive a constraint on $Z'$ from $Z\to \mu^+\mu^-Z'$ decay, as well as from the universality of $Z\to e^+e^-$ and $Z\to \mu^+\mu^-$~\cite{Ma:2001md}, as collectively shown by the red curves in Fig.~\ref{fig:exist} bottom panel.

When the $Z'$ couples to electrons, LEP-2 measurements of mono-photon events associated with large missing transverse energy at the $Z$ pole~\cite{DELPHI:2003dlq,DELPHI:2008uka} can also be used to set stringent limits on the coupling of $Z'$ to neutrinos. 
We show the mono-photon limits recast from Ref.~\cite{Fox:2011fx} as the magenta shaded region.

\item\textbf{LHC:} 
There exist dedicated searches for the $Z^\prime$ boson in the context of the $U(1)_{L_\mu-L_\tau}$ model by both ATLAS and CMS collaborations~\cite{ATLAS:2022wrd,CMS:2018yxg} using the $Z'$ production from the final-state radiation of $\mu$ or $\tau$ leptons in the Drell-Yan process. This is shown by the salmon (grey) shaded region for ATLAS (CMS) in Fig.~\ref{fig:exist}, which is the most stringent limit to date in most of the searched mass ranges. The olive-dashed and brown-dashed curves are limit from $4\ell$ search~\cite{CMS:2017dzg} and $3\ell$ search~\cite{CMS:2017moi} recast in Ref.~\cite{Drees:2018hhs}.

A search based on $Z'\to\mu^+\mu^-$ has been performed by the LHCb collaboration (similar searches were also done by BABAR~\cite{BaBar:2016sci} and Belle~\cite{Belle:2021feg}), which applies to the $U(1)_{L_e-L_\mu}$ and $U(1)_{L_\mu-L_\tau}$ models~\cite{LHCb:2019vmc}. Similarly, Belle II has performed searches for the $Z'$ boson using invisible $Z'$ decays in $e^+e^-\to \mu^+\mu^-Z'$~\cite{Belle-II:2022yaw} and also using visible $Z'$ decay to $\tau^+\tau^-$ resonance in $e^+e^-\to \mu^+\mu^-\tau^+\tau^-$~\cite{Belle-II:2023ydz}. These limits are applicable to the $U(1)_{L_\mu-L_\tau}$ case, but only for $M_{Z'}\lesssim 10$ GeV, and hence, are not shown here. 

Measurement of BR($Z\to \mu^+ \mu^-\tau^+ \tau^-$) by the CMS collaboration~\cite{CMS:2023nkg} will also be relevant for the $U(1)_{L_\mu-L_\tau}$ model; however, no limit on the $Z'$ contribution has been set by this analysis. We have also checked that the constraint on the $\gamma/Z$-$Z'$ kinetic mixing induced by the lepton loops~\cite{Cheung:2009qd,Bauer:2022nwt} from the direct $pp\to\ell^+\ell^-(\gamma)$ searches at the LHC~\cite{Hosseini:2022urq} are rather weak and do not show up in the range of our interest in Fig.~\ref{fig:exist}.

\end{itemize}

 Even with so many existing studies of the $U(1)_{L_\alpha-L_\beta}$ gauge boson as listed above, a large parameter space is allowed around the electroweak scale and remains to be explored at future colliders, shown by the blank region in Fig.~\ref{fig:exist}. We capitalize on this opportunity, and focus on the direct and indirect searches of the $Z'$ boson at future lepton colliders to extend the sensitivity coverage to higher masses and/or smaller couplings.

\section{$Z'$ phenomenology at future lepton colliders}
\label{sec:pheno}
In this section, we will explore the details of the $Z'$-boson phenomenology in the $U(1)_{L_\alpha-L_\beta}$ gauge model at future $e^+e^-$ and $\mu^+\mu^-$ colliders. An earlier study of the $U(1)_{L_\mu-L_\tau}$ scenario exists for a $\sqrt s=3$ TeV muon collider~\cite{Huang:2021nkl}; see also Refs.~\cite{Das:2022mmh, Sun:2023rsb, Jana:2023ogd} for related recent works. More careful considerations of the SM backgrounds, especially from the vector-boson fusion (VBF), will be examined in this work. In addition, we will extend our study to the $U(1)_{L_{e}-L_{\mu}}$ and $U(1)_{L_{e}-L_{\tau}}$ scenarios at both muon and electron-positron colliders. For concreteness and fair comparison, we will fix the center-of-mass energy at $\sqrt s=3$ TeV for both electron~\cite{CLICdp:2018cto} and muon~\cite{MuonCollider:2022xlm} collider options, unless otherwise specified.

%%%%%%%%%%%%%%%%%%%%%%%%%%%
\subsection{The $Z'$ resonance production} \label{subsec:resonance}
%%%%%%%%%%%%%%%%%%%%%%%%%%%
\begin{figure}
    \centering
    \subfigure[]{\includegraphics[width=0.24\textwidth]{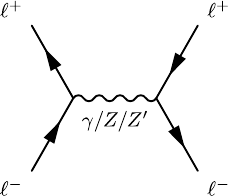}
    \label{fig:Feynll_a}
    }
    \subfigure[]{\includegraphics[width=0.24\textwidth]{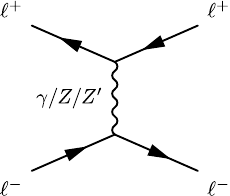}
    \label{fig:Feynll_b}
    }\\
    \subfigure[]{\includegraphics[width=0.23\textwidth]{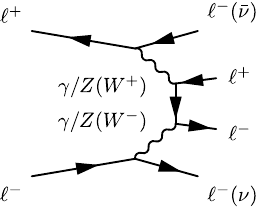}
    \label{fig:Feynll_c}
    }
    \subfigure[]{\includegraphics[width=0.23\textwidth]{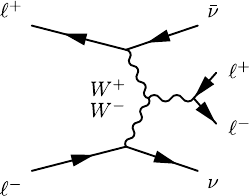}
    \label{fig:Feynll_d}
    }
    \subfigure[]{\includegraphics[width=0.23\textwidth]{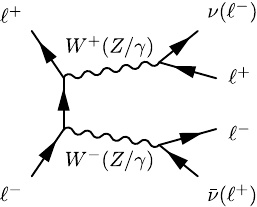}
     \label{fig:Feynll_e}
    }
    \subfigure[]{\includegraphics[width=0.23\textwidth]{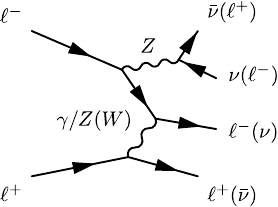}
     \label{fig:Feynll_f}
    }    
    \caption{Feynman diagrams for the lepton pair $\ell^+\ell^-$ production in the $U(1)_{L_\alpha-L_\beta}$ model (top row) and the representative vector-boson fusion or related backgrounds (bottom row) at high-energy lepton colliders.}
    \label{fig:Feynll}
\end{figure}

A pronounced signal for $Z'$ boson at a high-energy lepton collider can come from the direct lepton-pair annihilation, as shown in Fig.~\ref{fig:Feynll_a}.
Due to its short-lived nature in the parameter space of our interest [cf.~Fig.~\ref{fig:decay}], the $Z'$ will decay promptly before entering the detector. In the beam-lepton decay channel, we have extra $t$-channel diagrams with $Z'$ exchange, similar to the Bhabha scattering, as shown in Fig.~\ref{fig:Feynll_b}.
Two types of SM background will contribute to these dilepton final states. The first type comes from the same mechanism with annihilation or $t$-channel exchange but with the SM photon or $Z$ boson. The second type comes from the lepton pair production through 2-to-4 processes, $\ell^+\ell^-\to\ell'^+\ell'^-+\ell^+\ell^-(\nu\bar{\nu})$, in which the two forward/backward leptons are undetected, mainly induced by the neutral or charged current (NC or CC) VBF, as shown in  Figs.~\ref{fig:Feynll_c} and \ref{fig:Feynll_d}. They can fake the signal for the case with large initial state radiation (ISR). We note that, the diboson production in Fig.~\ref{fig:Feynll_e} and the three-body production $\ell^+\ell^- Z$ in Fig.~\ref{fig:Feynll_f},  
as well as many other crossing diagrams with $\ell^+\ell^- +  \nu\bar{\nu}$ final states, though potentially sizable, have rather different kinematics from the signal and can be effectively separated out, as we will comment on later. 

The characteristic feature of the resonance signal is the invariant mass peak obtained via the single $s$-channel cross-section
\begin{equation}
\sigma(s, M_{Z'})=\frac{g'^4}{12\pi}\frac{s}{(s-M_{Z'}^2)^2+M_{Z'}^2\Gamma_{Z'}^2} \, .
\end{equation}
At the peak $s=M_{Z'}^2$,  ignoring the interference and phase space acceptance, the rate is dominated by 
\begin{equation}
\sigma(\sqrt s = M_{Z'})= 
\frac{12\pi\ {\rm Br}_i\ {\rm Br}_f}{M_{Z'}^2},
\end{equation}
where $\textrm{Br}_{i(f)}$ is the $Z'$ decay branching fraction to the initial (final) state $i~(f)$. In reality, the annihilation energy (the ``partonic'' collision energy  $\sqrt{\hat{s}}$) is different from the designed collider energy ($\sqrt s$) due to the beam energy spread, and is typically lower mostly because of the ISR. Assuming a parton distribution function $f_\ell(x)$ for the contributing leptons with an energy fraction $x$, the observed cross-section at a given collider energy $\sqrt s> M_{Z'}$ is 
\begin{equation}
\sigma(\sqrt s)= \int_0^1 \dd x \: f_\ell(x) 
\sigma(s, M_{Z'}) \approx
12\pi^2\ {\rm Br}_i\ {\rm Br}_f\ 
\frac{\Gamma_{Z'}}{M_{Z'}^3} \  f_\ell\left({M_{Z'}^2/s}\right) ,
\end{equation}
where the narrow-width approximation (NWA) has been adopted with the on-shell condition
\begin{equation}\label{eq:momZp}
\hat{s}=xs\approx M^2_{Z'}.
\end{equation}

\begin{figure}[tb]
\centering
\includegraphics[width=0.49\textwidth]{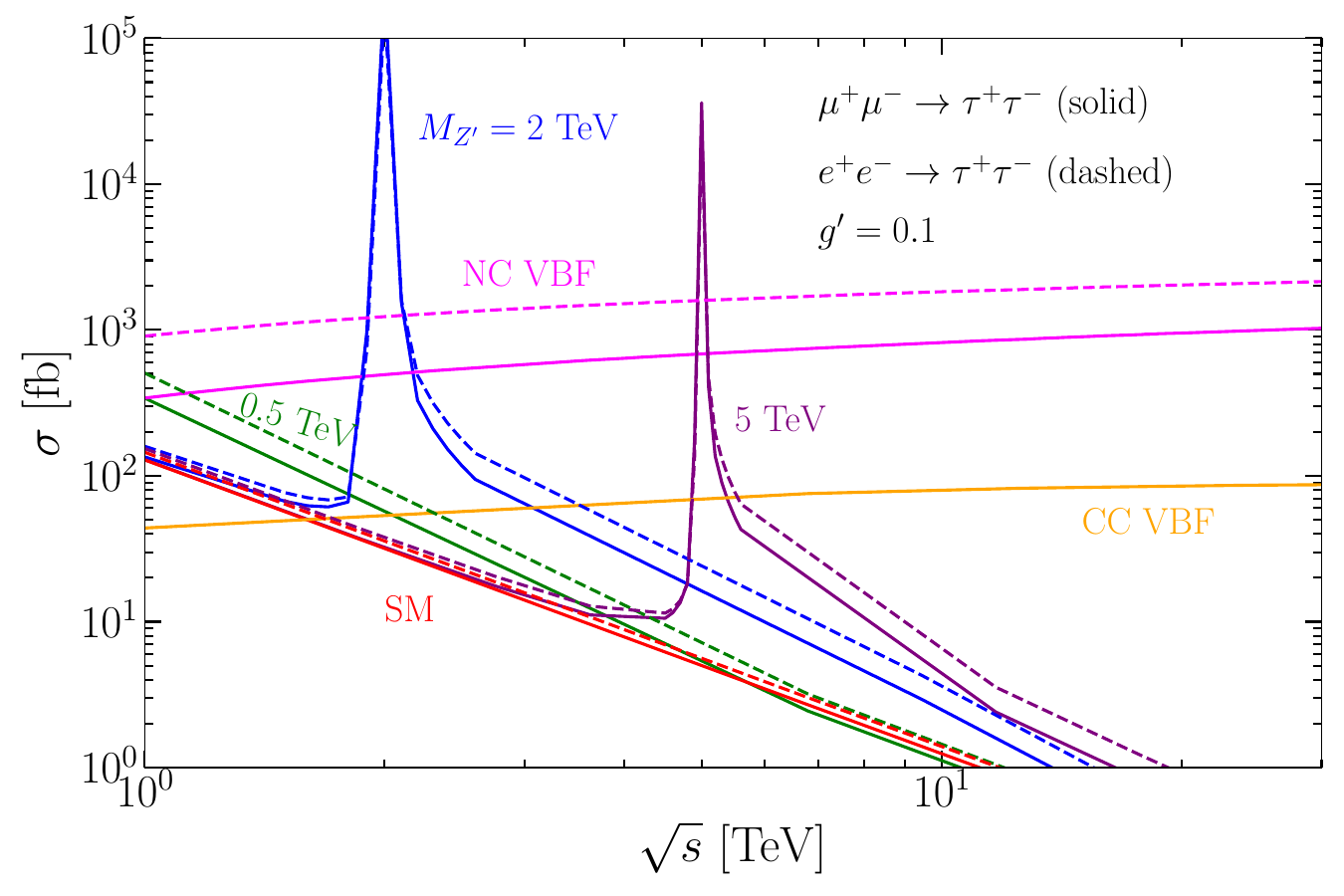}
\includegraphics[width=0.49\textwidth]{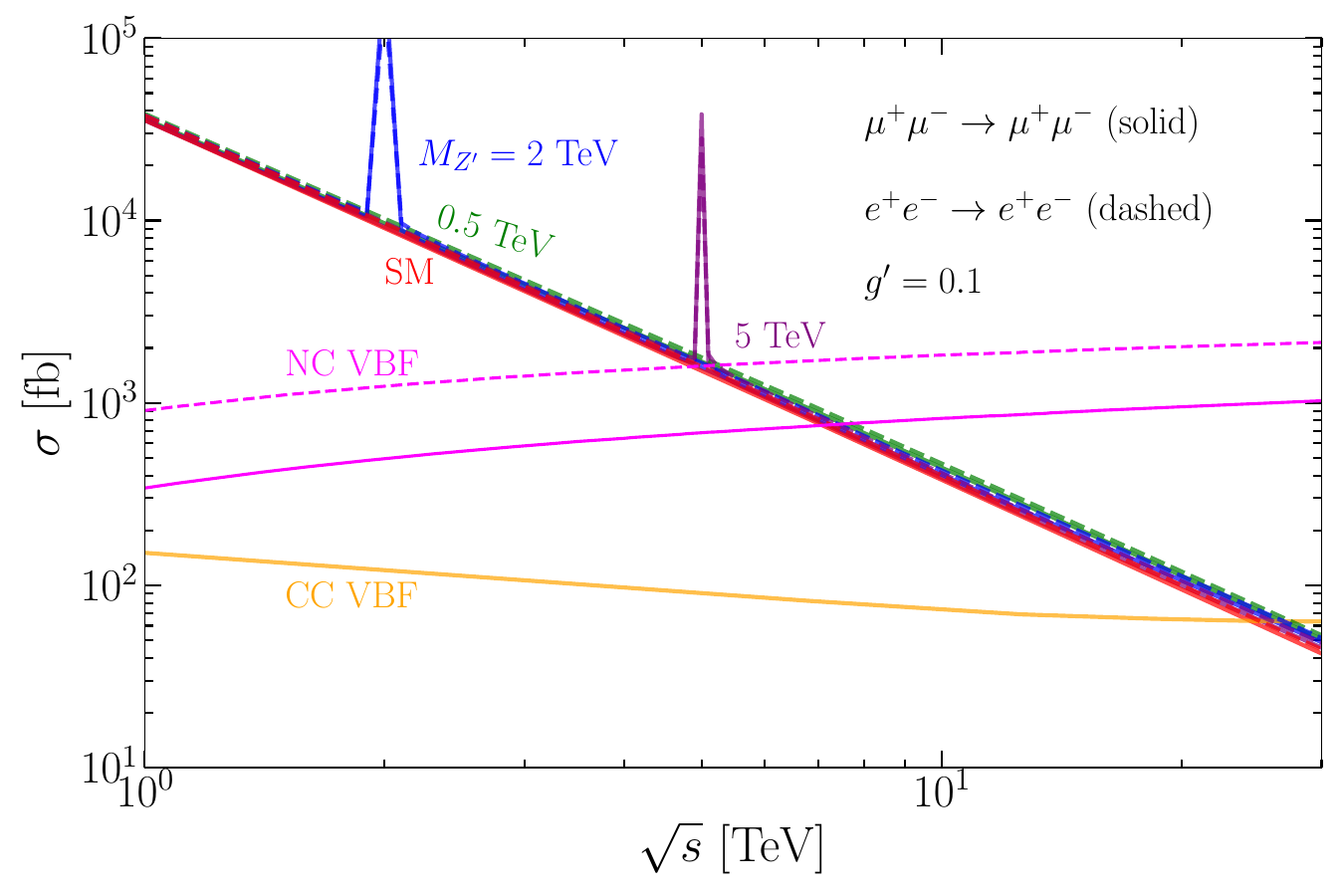}
\caption{Lepton-pair production cross-section versus the electron/muon collider energy in the SM and $U(1)_{L_\alpha-L_\beta}$ models. We take the $\tau^+\tau^-$ production as a representative for processes with final-state leptons different from initial beams on the left panel, while the one with the same initial- and final-state lepton flavors is shown on the right panel.
The pre-selection cuts in Eq.~(\ref{eq:PSC}) have been applied here. 
}
\label{fig:Ecm}
\end{figure}

In Fig.~\ref{fig:Ecm}, we present the signal and background cross-sections for the lepton-pair production at high-energy electron and muon colliders in both the SM and the $U(1)_{L_\alpha-L_\beta}$ extended model, including the ISR effects for the resonance production. 
In the $U(1)_{L_\alpha-L_\beta}$ model, we fix $g'=0.1$ and $M_{Z'}=0.5/2/5~\TeV$ to demonstrate the features when the $Z'$ mass is fully below the collider energy $\sqrt{s}$ and when it is being crossed as $\sqrt{s}$ increases.
Here, on the left panel, we take the $e^+e^-/\mu^+\mu^-\to\tau^+\tau^-$ as representatives for the processes with final-state leptons different from initial-beam flavors, which probe the $U(1)_{L_e-L_\tau}$ and $U(1)_{L_\mu-L_\tau}$ models, respectively. The $U(1)_{L_e-L_\mu}$ can be probed with either $e^+e^-\to\mu^+\mu^-$ or $\mu^+\mu^-\to e^+e^-$ scatterings, which show similar behaviors as $e^+e^-/\mu^+\mu^-\to\tau^+\tau^-$, or via the Bhabha scatterings $e^+e^-\to e^+e^-$ and $\mu^+\mu^-\to \mu^+\mu^-$, as shown on the right panel.
In the rest of our simulation, we have imposed the universal pre-selection cuts (PSCs) 
\begin{equation}~\label{eq:PSC}
p_T^{\ell}>30~\GeV, ~|\eta_{\ell}|<2.44,~\Delta R_{\ell\ell}>0.3
\end{equation}
for final-state leptons, which are essential to regulate collinear divergence in the Bhabha scattering and to simulate the detector acceptance. 
The lepton pseudo-rapidity cut $|\eta_{\ell}|<2.44$ corresponds to the detector coverage within $10^{\degree}<\theta<170^{\degree}$.
The signal cross-sections are calculated using WHIZARD~\cite{Moretti:2001zz,Kilian:2007gr} and cross-checked using MadGraph~\cite{Alwall:2014hca}, after interfacing with the UFO model files generated using FeynRules~\cite{Christensen:2010wz}. The QED ISRs are treated with WHIZARD, which resums soft photons to all orders and hard-collinear ones up to the third order~\cite{Kilian:2007gr,Moretti:2001zz}.

In Fig.~\ref{fig:Ecm}, we see two types of behavior of the curves. The downward-going curves with the $1/s$ behavior correspond to the annihilation or regulated $t$-channel processes. The cross-section difference for the SM annihilation processes between the two panels is from the additional $t$-channel contributions in the same initial-final flavor case, $\ell^+\ell^-\to\ell^+\ell^-~(\ell=e,\mu)$. When $\sqrt{s}\sim M_{Z'}$, we see a big enhancement in the lepton-pair cross-section due to the resonant production. 
Off the resonance when $\sqrt{s} > M_{Z'}$, there is still an enhancement peaked at $M(\ell^+\ell^-)\sim M_{Z'}$, due to the ISR, namely, the ``radiative return''~\cite{Denig:2006kj}. For $M_{Z'} \gg \sqrt{s}$ on the other hand, the first two diagrams in Fig.~\ref{fig:Feynll} both contribute to the signal, with the cross-section scaling as $\sigma \sim 1/M_{Z'}^2$.

In contrast, the upward-going curves in Fig.~\ref{fig:Ecm} indicate the NC and CC VBF processes, mediated by $\gamma/Z$ and $W$ bosons, respectively. The VBF backgrounds for both NC and CC are calculated with the WHIZARD fixed-order (FO) calculation. 
Since the photon-photon initiated processes dominate the NC contribution, 
we verified the above calculations with MadGraph's equivalent photon approximation (EPA)~\cite{Budnev:1975poe}, {\it i.e.}, improved Weizsacker-Williams approximation~\cite{Frixione:1993yw}.
The VBF channels serve as a significant part of the SM background, in particular in the off-resonance region. 
With the collider energy at $\sqrt{s}=3~\TeV$ of our interest, we see the VBF cross-section can even dominate the lepton-pair production, in both electron and muon collider scenarios. With respect to the NC VBF cross-sections for electron colliders, the muon ones are generally smaller, due to their large mass, which reduces the photon radiations. A similar situation occurs in the ISR annihilation cross-section as well. In comparison, the CC VBF cross-sections for the lepton pair production at electron and muon colliders are largely the same, as both lepton masses are negligible with respect to the $W$-boson one. We have also included the Higgs decay $H\to \tau\tau$, which yields about 27 fb at 30 TeV. 
The notable larger cross-section at lower energies in the second panel than in the first for CC VBF is owing to additional channels for the same initial- and final-state flavor, as well as the three-body contribution $\ell^+\ell^- \to \ell^+ \ell^- Z$ in Fig.~\ref{fig:Feynll_f}. 
As to be shown with optimization cuts later, these are only appreciable near the threshold for a heavy $Z'$. Although the backgrounds from 2-body [cf.~Fig.~\ref{fig:Feynll_e}] and 3-body [cf. Fig.~\ref{fig:Feynll_f}] processes are sizable, they still fall below the SM $s$-channel contribution. Furthermore, they could be effectively separated from the signal by examining the large $p_T(\ell^+\ell^-)$ as opposed to $p_T(\ell^+\ell^-)\approx 0$ for the signal.

\begin{figure}[tb]
\centering
\includegraphics[width=0.49\textwidth]{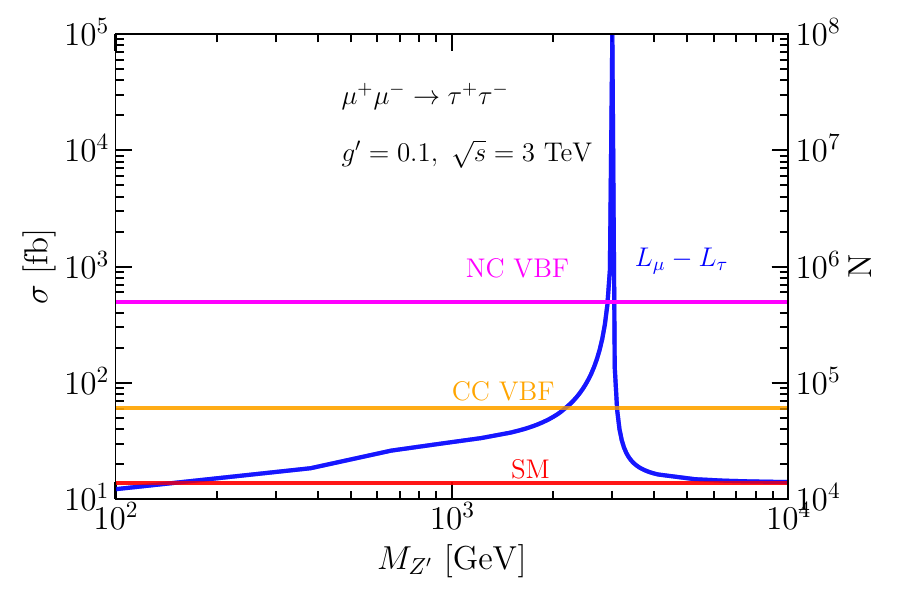} 
\includegraphics[width=0.49\textwidth]{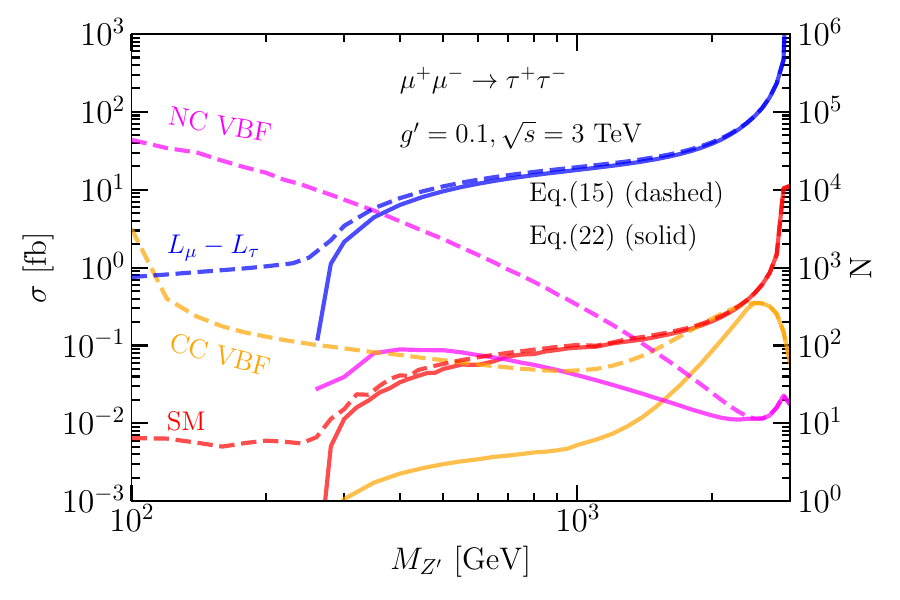}
\caption{Lepton-pair annihilation cross-sections versus the new gauge boson mass $M_{Z'}$ at a $\sqrt{s}=3~\TeV$ muon collider. {\it Left:} With PSCs in Eq.~(\ref{eq:PSC}).
{\it Right:} With the optimization cuts $|M_{\tau\tau}-M_{Z'}|<0.05M_{Z'}$ (dashed curves), and in addition, $|y\pm y_{Z'}|<0.2$ (solid curves). 
We also include the number of events on the right y-axis, which corresponds to an integrated luminosity $\mathcal{L}=1~\textrm{ab}^{-1}$. 
}
\label{fig:xsec3TeV}
\end{figure}

In the left panel of Fig.~\ref{fig:xsec3TeV}, we present the lepton-pair annihilation cross-sections versus the gauge boson mass $M_{Z'}$ in the $U(1)_{L_\alpha-L_\beta}$ model, with gauge coupling fixed as $g'=0.1$, with the same cuts as in Fig.~\ref{fig:Ecm}. 
We see that when $M_{Z'} < \sqrt{s}$, the cross-section increases while  approaching the resonant energy. In contrast, for $M_{Z'} > \sqrt{s}$, we see the $U(1)_{L_\alpha-L_\beta}$ cross-section asymptotically approaches to the SM one, scaled as $\sigma \sim 1/M_{Z'}^2$, suggesting the limitation of the direct probe of heavy $Z'$ at a lepton collider.
In the right panel, we show the asymptotic behavior of the cross-sections when $\sqrt{s} \to M_{Z'}$, with the optimal cuts as labeled on the plot. 
In both panels, we also include the SM background cross-sections from the direct annihilation, as well as from the NC and CC VBF processes, for comparison.

\begin{figure}
\centering
\includegraphics[width=0.45\textwidth]{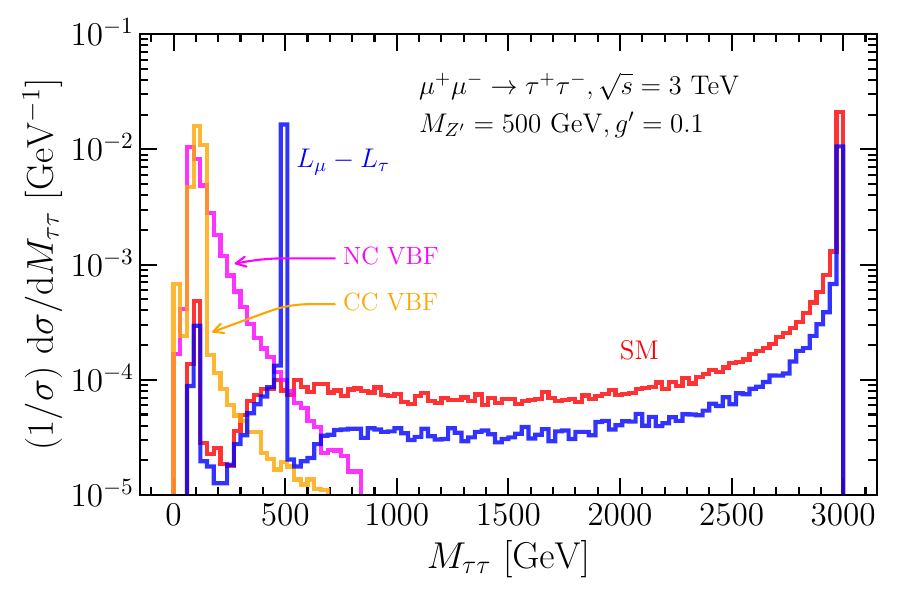}
\includegraphics[width=0.45\textwidth]{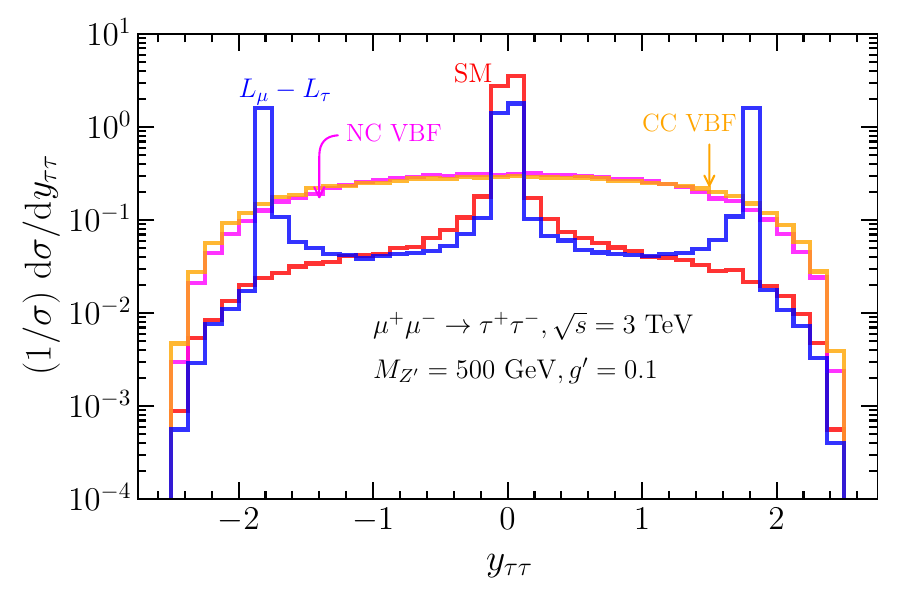}    
\includegraphics[width=0.45\textwidth]{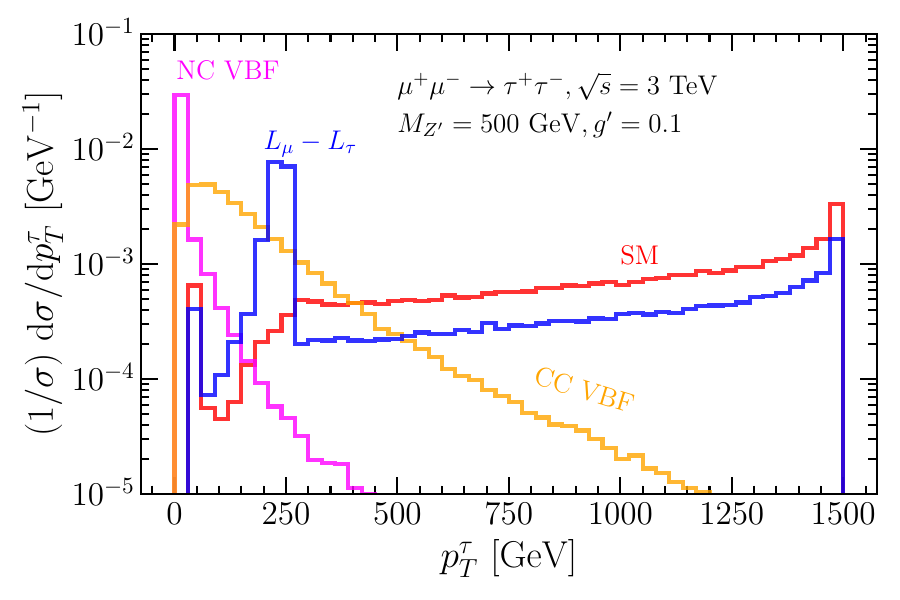}
\includegraphics[width=0.45\textwidth]{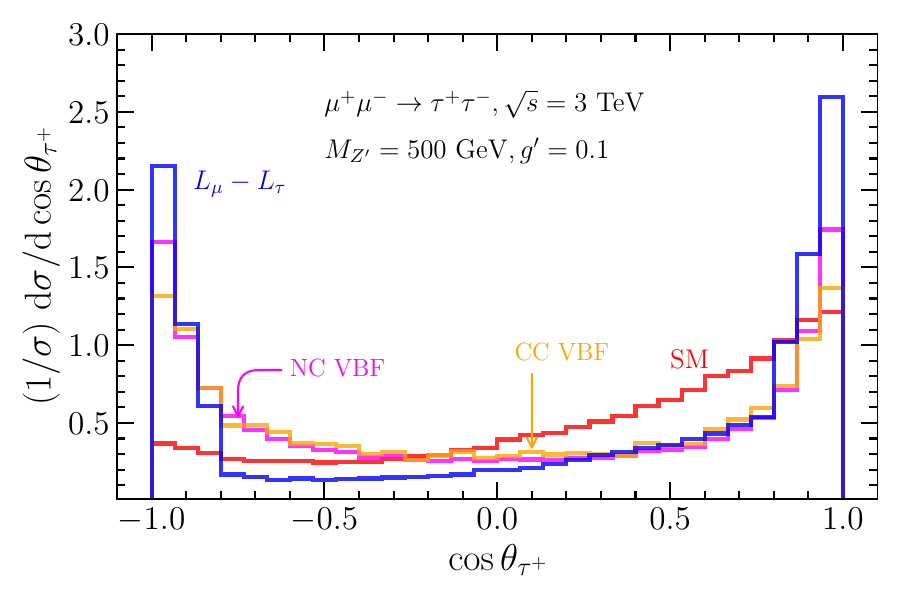}
\caption{Normalized kinematic distributions for the invariant mass $M_{\tau\tau}$ and rapidity $y_{\tau\tau}$ of the lepton pair, as well as the transverse momentum $p_T^{\tau^+}$ and angle cosine $\cos\theta_{\tau^+}$ of the positively-charged final-state $\tau$ lepton in the $\mu^+\mu^-\to\tau^+\tau^-$ scattering at a $\sqrt s=3$ TeV muon collider. Here we have fixed $M_{Z'}=500$ GeV and $g'=0.1$ for the signal.
}
\label{fig:dist_mumu2tata}
\end{figure}

We take $\mu^+\mu^-\to\tau^+\tau^-$ process as an example to probe the $U(1)_{L_\mu-L_\tau}$ model. 
Even with potentially larger SM background cross-sections, we can still hope to separate the $Z'$ signal from the backgrounds based on their kinematic features. 
%The most characteristic 
The normalized distributions of invariant mass $M_{\tau\tau}$ and rapidity $y_{\tau\tau}$ for the final-state lepton pairs, as well as the transverse momentum $p_T^{\tau^+}$ and cosine angle $\cos\theta_{\tau^+}$ for the positively-charged final-state lepton with respect to the $e^+/\mu^+$ direction are shown in Fig.~\ref{fig:dist_mumu2tata} for both signal and backgrounds, with PSCs in Eq.~(\ref{eq:PSC}).
For the $U(1)$ gauge model, we take $M_{Z'}=500~\GeV$ and $g'=0.1$ for demonstration. 

\begin{table}
\centering
\begin{tabular}{c|c|c|c||c|c|c}
\hline\hline
$U(1)_{L_\mu-L_\tau}$ & \multicolumn{3}{c||}{Signal} & \multicolumn{3}{c}{SM backgrounds}\\
\hline
$M_{Z'}~[\TeV]$ & 0.5 & 2 & 5 & Annihilation & CC VBF & NC VBF\\
\hline\hline
$\sigma$ [fb] & \multicolumn{5}{c}{$\mu^+\mu^-\to\tau^+\tau^-$} \\
\hline\hline
%$|\eta_\ell|<2.44$ & 22.6 &  51.1& 15.2 & 14.0 & 95.9 & 1.66e5 \\
Eq.~(\ref{eq:PSC}) & 22.5 & 50.9 & 15.0  &13.8 & 61.3 & 510\\
\hline
$0.475<M_{\ell\ell}/\TeV<0.525$ & 11.1 & &  & $5.86\cdot10^{-2}$ & $6.48\cdot10^{-2}$  & 2.31 \\
$1.59<|y_{\tau\tau}|<1.99$ & 9.6 &  & & $5.07\cdot10^{-2}$ & $3.00\cdot10^{-3}$  & $8.78\cdot10^{-2}$ \\
\hline
$1.9<M_{\ell\ell}/\TeV<2.1$ & & 40.7 & & 0.214  & $2.10\cdot10^{-1}$ & $2.49\cdot10^{-2}$ \\
$0.20<|y_{\tau\tau}|<0.60$ &  & 39.1&  & 0.205 & $8.30\cdot10^{-2}$  &$1.30\cdot10^{-2}$ \\
\hline
$M_{\ell\ell}>0.95\sqrt{s}$ & &  & 11.0 & 11.0& $6.05\cdot10^{-2}$ & $1.76\cdot10^{-2}$  \\
\hline\hline
\end{tabular}
\caption{The cut-flow table $\tau^+\tau^-$ pair production at a $\sqrt{s}=3~\TeV$ muon collider. 
The gauge coupling is fixed at $g'=0.1$.}
\label{tab:cutflow}
\end{table}

For the $2\to 2$ annihilation processes in the SM, the invariant mass is primarily peaked around the collision energy $M_{\tau\tau}=\sqrt{s}$, with a long tail at lower energies from the ISR. 
In contrast, for the $Z'$ signal, a resonance peak shows up at  $M_{\tau\tau}=M_{Z'}<\sqrt{s}$
as a result of the so-called ``radiative return'', indicating the potential to discover this new $Z'$ particle, with the signal rate scaled as the coupling strength $g'^2$. 
In comparison with the annihilation case, we see the distributions of the VBF channels, including both NC and CC, die out very quickly at the high invariant mass, as the parton luminosity decreases as $1/M_{\tau\tau}^2$. As such, the SM VBF backgrounds would possess less of a problem for a large $M_{Z'}$. 
We adopt an invariant mass selection cut to optimize the search for $M_{Z'}$ as 
\begin{equation}\label{eq:mzcut}
\begin{aligned}
&|M_{\ell\ell}-M_{Z'}|< 0.05 M_{Z'}~(10~\GeV) 
\ {\rm \ for\ resonance}\ 
M_{Z'} < \sqrt{s}\, ;\\ 
&M_{\ell\ell}>0.95\sqrt{s} 
\ {\rm for\ off~shell}\ 
M_{Z'}\geq\sqrt{s}\, .
\end{aligned}
\end{equation}
for final-state tau (electron or muon) leptons, respectively.
Here, the final-state tau requires a reconstruction from the tau decay products, in which case we consider a looser mass window as $0.05M_{Z'}$, while the final-state electrons and muons can be well observed in the detector, in which we take a 10 GeV mass window~\cite{CLICdp:2018cto}.

Another characteristic feature comes from the observation that the dominant configuration in the radiative return is from a leading single photon radiation. The kinematics can be determined through the 2-to-2 scattering process
\begin{equation}
\mu^+(p_1)+\mu^-(p_2)\to Z'(p_{Z'})+\gamma(p_\gamma)\, .
\end{equation}
The final-state momenta can be parameterized in terms of the photon transverse momentum $p_T$ and pseudo-rapidity $\eta$ as
\begin{equation}
\begin{aligned}
p_\gamma&=(p_T\cosh\eta,\vec{p}_T,p_T\sinh\eta)\, ,\\
p_{Z'}&=(\sqrt{s}-p_T\cosh\eta,-\vec{p}_T,-p_T\sinh\eta)\, ,
\end{aligned}
\end{equation}
where we have employed the momentum conservation $p_1+p_2=p_{Z'}+p_\gamma=(\sqrt{s},\vec{0},0)$. With the on-shell condition $p_{Z'}^2=M_{Z'}^2$, we obtain the photon transverse momentum as
\begin{equation}
 p_T=\frac{s-M_{Z'}^2}{2\sqrt{s}\cosh\eta}\, .   
\end{equation}
The final-state $Z'$-boson rapidity can be analytically determined as
\begin{equation}
\label{eq:yZp}
y(\eta)=\frac{1}{2}\log\frac{p_{Z'}^0+p_{Z'}^{3}}{p_{Z'}^0-p_{Z'}^3}
=\frac{1}{2}\log\frac{se^\eta+M_{Z'}^2e^{-\eta}}{se^{-\eta}+M_{Z'}^2e^\eta}\, .
\end{equation}
In the spirit of ``radiative return'' with a collinear photon, we have $|\eta|\to\infty$ ({\it i.e.}, $p_T\to0)$, the $Z'$ rapidity becomes
\begin{equation}
|y(\pm\infty)|=y_{Z'}\equiv\log(\sqrt{s}/M_{Z'}).
\end{equation}
In this case, the momentum fraction carried by the photon becomes
\begin{equation}
\bar{x}\equiv1-x=\frac{p_T\cosh\eta}{\sqrt{s}/2}\xrightarrow{|\eta|\to\infty}1-\frac{M_{Z'}^2}{s}  \, ,   
\end{equation}
while the beam lepton carries a fraction $x=M_{Z'}^2/s$, corresponding to the on-shell condition in Eq.~(\ref{eq:momZp}). This is remarkable since it predicts the mono-chromatic value of the rapidity of $Z'$ for a given mass at a fixed collider energy, which would single out the resonant signal over the continuum SM background. 
Examining the lepton-pair rapidity distribution in Fig.~\ref{fig:dist_mumu2tata}, we see in the annihilation processes that the 500 GeV $Z'$ signal peaks at 
$|y_{\tau\tau}|= \log(\sqrt{s}/M_{Z'}) \approx 1.79$, whereas the SM contribution is primarily peaked around $y_{\tau\tau}\sim0$, and  the VBF processes are spread out. 
This motivates us to impose a rapidity selection cut, for a given hypothetical $M_{Z'}$, 
\begin{equation}
    \left|y_{\tau\tau}\pm y_{Z'}\right|<0.2 \, ,
    \label{eq:y}
\end{equation} 
to increase the signal-to-background ratio further, which is shown on the right panel of Fig.~\ref{fig:xsec3TeV} by the solid curves.

However, there is an additional complication. For $M_{Z'} \ll \sqrt{s}$, $y_{Z'}\approx \log\cot(\theta/2)$, which results in \begin{equation}
\tan(\theta/2) \approx M_{Z'}/\sqrt{s}. 
\end{equation}
Assuming the minimal detector acceptance in the polar angle being about $10^\circ\ (|\eta_\ell|< 2.44),$
then a particle of a mass $M< 0.088 \sqrt s$ would mostly escape from the detection into the beam pipe, 
which leads to a missing particle of mass $M=(261, 872, 2615)$ GeV at a collider of $\sqrt s =(3, 10, 30)$ TeV. 
As a caveat,
this rapidity optimization cut in Eq.~(\ref{eq:y}) would not be applicable when $M_{Z'}<0.088\sqrt s$ as $y_{Z'}$ goes beyond the detector acceptance. 

In Table \ref{tab:cutflow}, we demonstrate our cut-flow strategy with $g'=0.1$ and $M_{Z'}=0.5,~2$ and 5 TeV as examples. We see that the cuts are highly effective in preserving the signal. The cut on $M_{\ell\ell}$ is effective for both annihilation and VBF backgrounds, while that on $y_{\tau\tau}$ is more on reducing VBF background, by orders of magnitude. 
The cross-sections for both signal and backgrounds with optimization cuts $|M_{\ell\ell}-M_{Z'}|<0.05M_{Z'}$ and subsequently $|y_{\ell\ell}\pm y_{Z'}|<0.2$ (which only applies when $M_Z'>261~\GeV$) at a  $\sqrt{s}=3~\TeV$ muon collider are also shown in Fig.~\ref{fig:xsec3TeV} (right panel).

\begin{figure}
    \centering
    \includegraphics[width=0.48\textwidth]{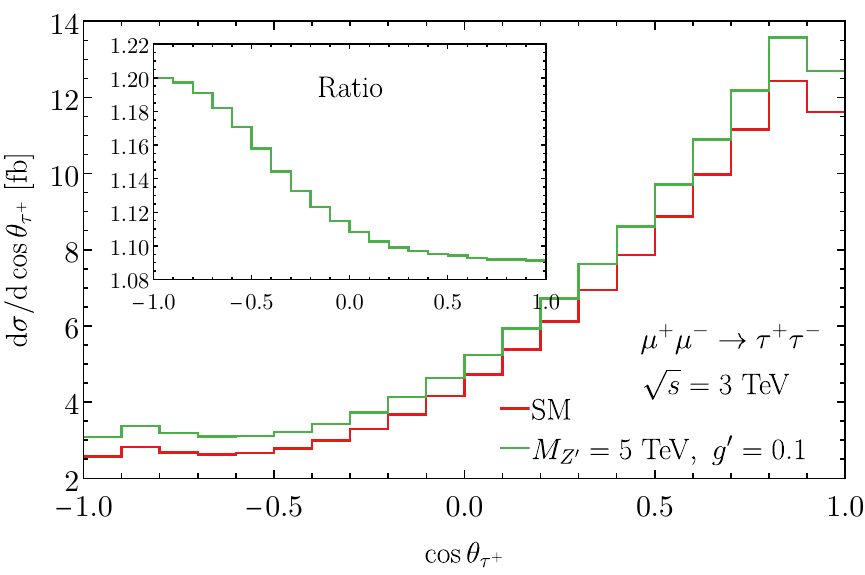}
    \includegraphics[width=0.51\textwidth]{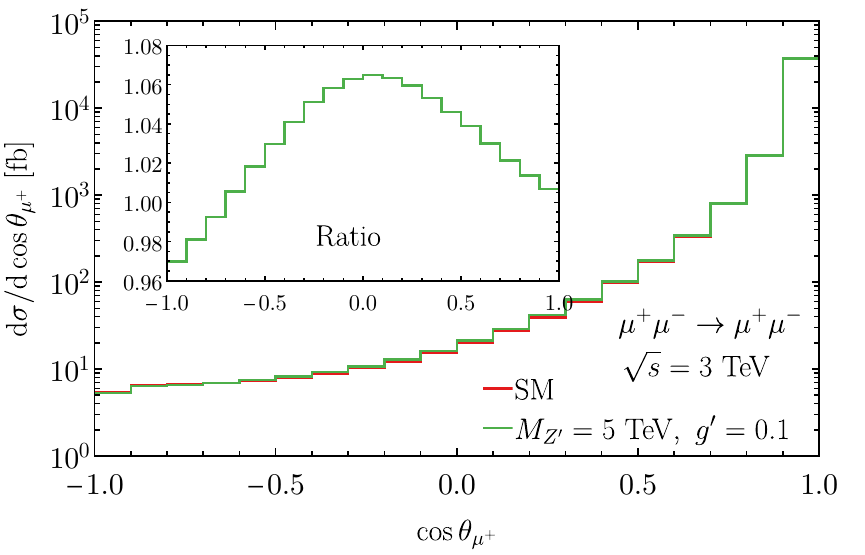}    
    \caption{The cosine angle distribution for the final-state leptons in the off-shell $s$-channel $\mu^+\mu^-\to\gamma/Z/Z'\to\tau^+\tau^-$ (left) and $s/t$-channel $\mu^+\mu^-\xrightarrow{\gamma/Z/Z'}\mu^+\mu^-$ (right) scatterings. 
    The $Z'$ signal comes from the difference between the SM and the $Z'$ model, with the relative size of $(S+B)/B$ shown in the corresponding inset.
    }
    \label{fig:costh}
\end{figure}

%%%%%%%%%%%%%%%%%%%%%%%%%%%%%%%%%% 
\subsection{Off-shell $Z'$ production}
\label{sec:offshell}
%%%%%%%%%%%%%%%%%%%%%%%%%%%%%%%%%%%
When $M_{Z'}>\sqrt{s}$, the $Z'$ resonance cannot be produced directly on-shell at a collider. However, the off-shell $Z'$-mediated diagrams, both in the $s$- and $t$-channels,  as shown in  Figs.~\ref{fig:Feynll_a} and~\ref{fig:Feynll_b} respectively, are expected to interfere with the SM ones, which yields a modification to the final-state lepton distribution with respect to the SM ones at the order of $s/M_{Z'}^2$. In such a way, an indirect sensitivity to the $Z'$ boson can be placed based on precision measurements, {\it e.g.}, the forward-backward asymmetry (FBA). 
In Fig.~\ref{fig:costh}, we show the cosine angle distributions of the final-state lepton in the $s$-channel $\mu^+\mu^-\to\gamma/Z/Z'\to\tau^+\tau^-$ and the $s/t$-channel $\mu^+\mu^-\xrightarrow{\gamma/Z/Z'}\mu^+\mu^-$ processes, with the benchmark values of $M_{Z'}=5~\TeV$ and $g'=0.1$, and with the same pre-selection cuts as in Eq.~(\ref{eq:PSC}). We see that in the $s$-channel scattering $\mu^+\mu^-\to\tau^+\tau^-$, the additional $Z'$ mediated process not only changes the shape of angle distribution, but also enhances the total rate. In the Bhabha-like scattering $\mu^+\mu^-\to\mu^+\mu^-$, the total rate largely remains unchanged, mainly due to the dominant $t$-channel $\gamma$-mediated background, while the FBA gets modified. 
Later, we will perform a bin-by-bin angular distribution analysis in both cases, which implicitly includes the FBA information.

 %%%%%%%%%%%%%%%%%%%%%%%%%%%
\subsection{$Z'+\gamma$ associated production} \label{subsec:zpgama}
%%%%%%%%%%%%%%%%%%%%%%%%%%%%%%%%%%
\begin{figure}
    \centering
    \includegraphics{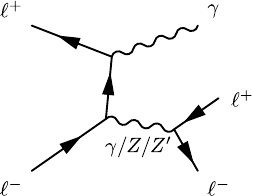}
    \includegraphics{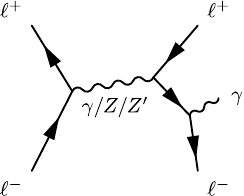}    
    \caption{Representative Feynman diagrams for lepton-pair production associated with a photon at a lepton  collider.}
    \label{fig:FeynllA}
\end{figure}

In the ISR for the resonance production as explored above, the soft or collinear photons are unobservable in the detector. In contrast, a radiated photon can be detected as long as it is within the acceptance of the electromagnetic calorimeter. 
We show the corresponding Feynman diagrams in Fig.~\ref{fig:FeynllA}, including channels with the same or different initial- and final-state lepton flavors. 
Different from the channels in Fig.~\ref{fig:Feynll}, the resolved photon here requires additional acceptance, which we choose as 
\begin{equation}\label{eq:PSC2}
p_{T}^{\gamma}>30~\GeV, ~ |\eta_{\gamma}|<2.44, ~\Delta R_{\gamma\ell}>0.3,    
\end{equation}
on top of the PSCs in Eq.~(\ref{eq:PSC}). Although the total signal rate will be smaller due to the additional hard photon radiation, the unique kinematics may help signal identification and property studies. 

\begin{figure}
\centering
\includegraphics[width=0.49\textwidth]{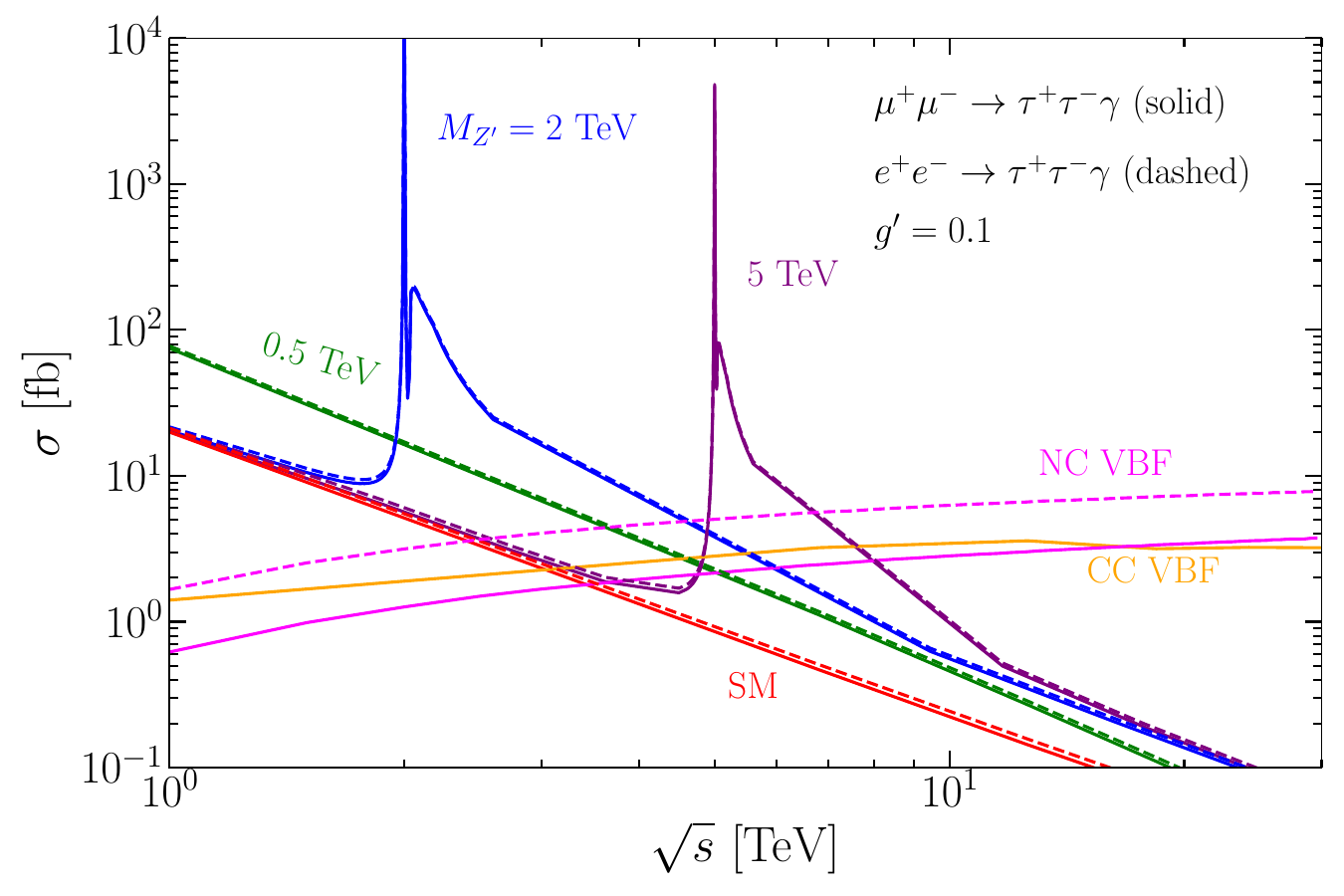}
\includegraphics[width=0.49\textwidth]{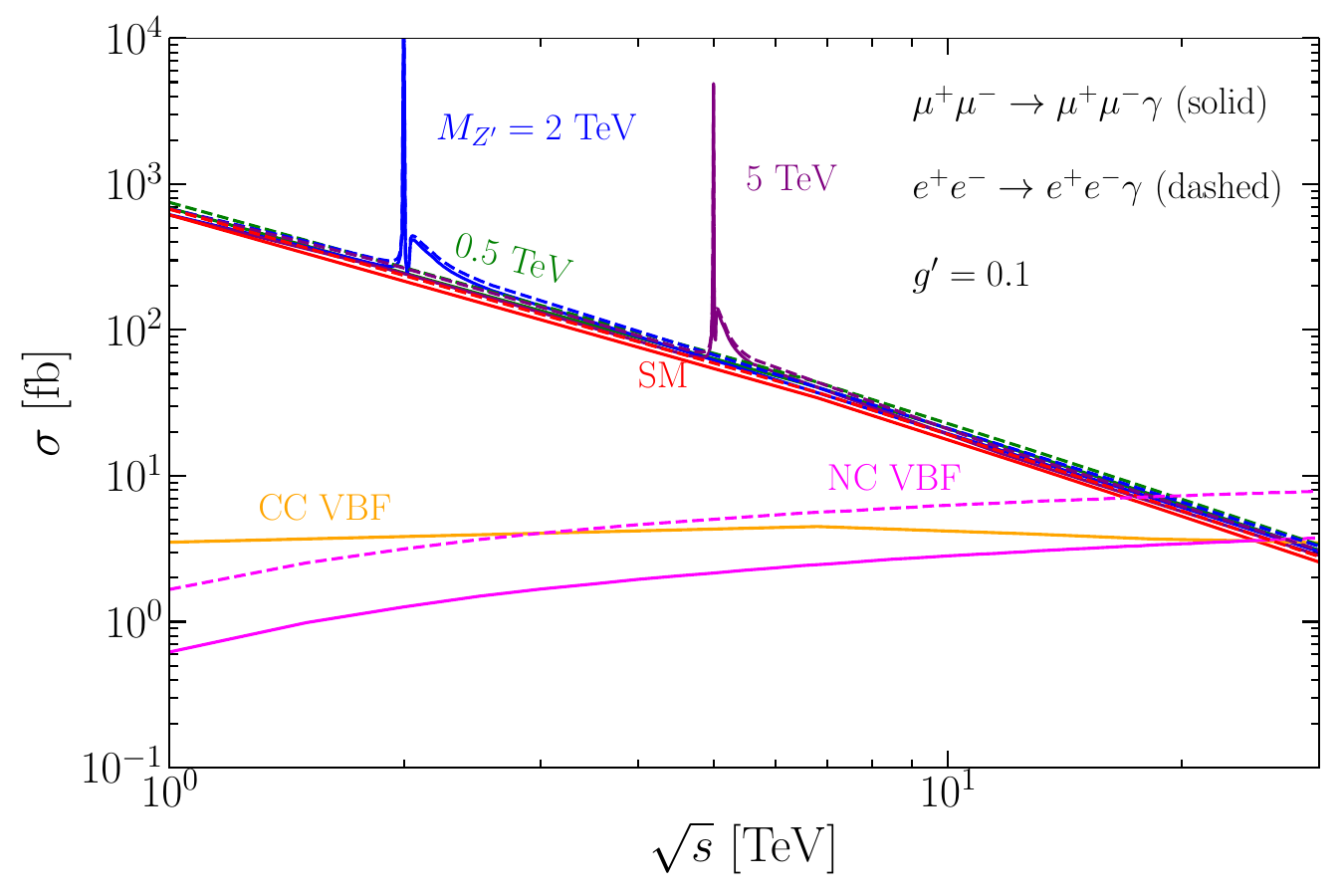}
\caption{The cross-sections for $e^+e^-/\mu^+\mu^-\to\tau^+\tau^-\gamma$ (left) and $e^+e^-(\mu^+\mu^-)\to e^+e^-(\mu^+\mu^-)\gamma$ (right). 
}
\label{fig:xsllA}
\end{figure}

In Fig.~\ref{fig:xsllA}, we show the cross-sections for the $Z'+\gamma$ production with both signal and SM backgrounds 
with the pre-selection cuts in Eq.~\eqref{eq:PSC} plus Eq.~(\ref{eq:PSC2}). 
As before, the signal and SM annihilation cross-sections are calculated with WHIZARD and cross-checked with MadGraph. The NC VBF is taken with the MadGraph EPA, while the CC VBF is done with the WHIZARD FO calculation. 
Similar to the direct $Z'$ production via ISR in the last section, there is a sharp peak around $\sqrt{s}\approx M_{Z'}$, as a result of the resonant production and decay of $Z' \to \ell^+\ell^- \gamma$.
We also get notable resonance enhancements when $\sqrt{s}\gtrsim M_{Z'}$, while the additional resolved photon requires a slightly larger collider energy $\sqrt s \sim E_{Z'} +E_\gamma$. 
Different from the direct production channel, we see the VBF cross-sections, as well as the other higher-order diagrams $W^+W^-\gamma,\ ZZ\gamma$,  are significantly suppressed, as a result of both one additional electric gauge coupling and the photon selection cuts. As a result, the SM background mainly comes from the diagrams in Fig.~\ref{fig:FeynllA} with a $\gamma/Z$ mediation. 

\begin{figure}[tb]
\centering
\includegraphics[width=0.49\textwidth]{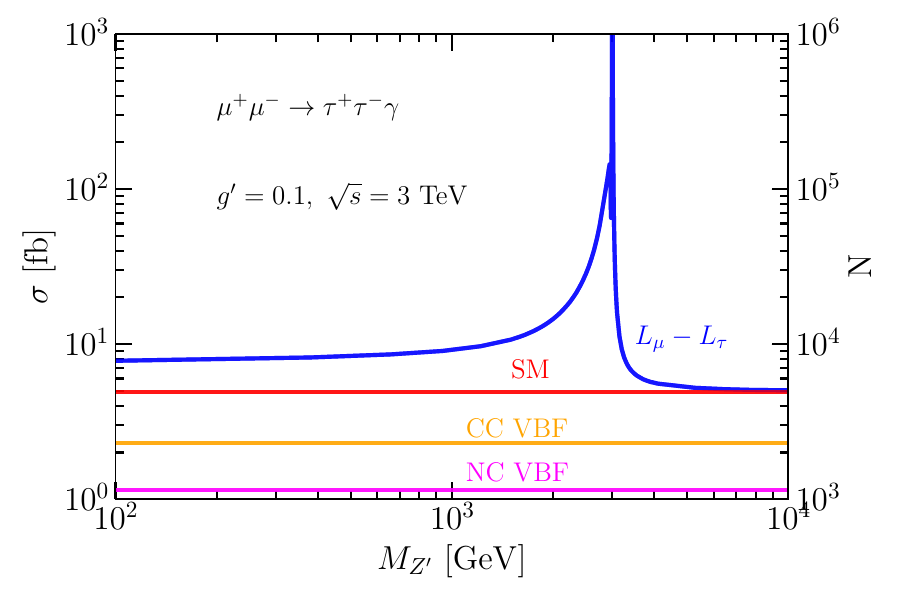}
\includegraphics[width=0.49\textwidth]{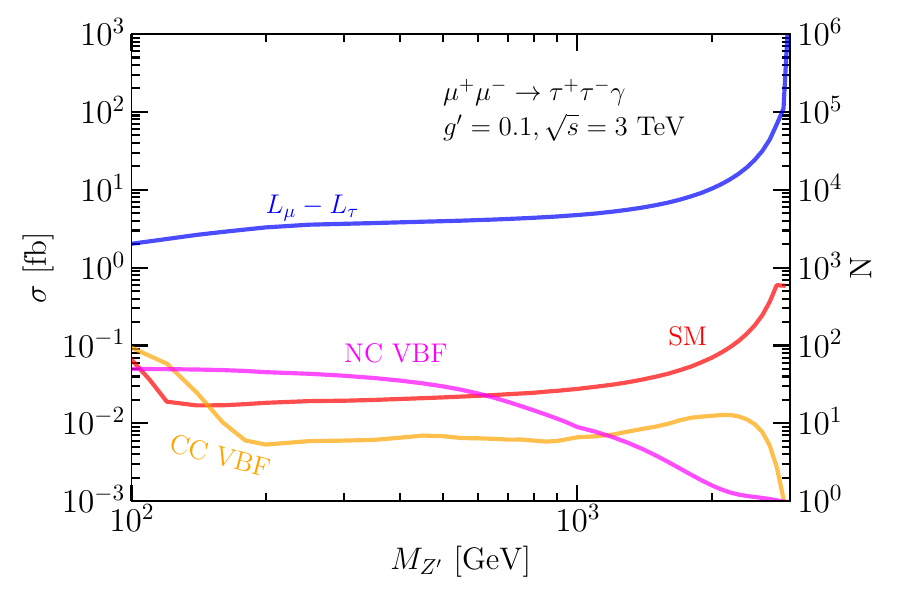}
\caption{{\it Left:} The pre-selection cross-section of $\tau\tau+\gamma$ associated production versus the $Z'$-boson mass $M_{Z'}$ at a 3 TeV muon collider. {\it Right:} The same cross-section, but with additional invariant mass cut $|M_{\tau\tau}-M_{Z'}|<0.05M_{Z'}$. }
\label{fig:llA3TeV}
\end{figure}

In Fig.~\ref{fig:llA3TeV} (left), we show the $Z'+\gamma$ cross-section at a 3 TeV muon collider as a function of the $Z'$ mass, with the gauge coupling fixed at $g'=0.1$. We have used the PSCs in Eq.~(\ref{eq:PSC}) together with Eq.~(\ref{eq:PSC2}). 
In comparison with the single $Z'$ production in Fig.~\ref{fig:xsec3TeV}, the $Z'+\gamma$ cross-section gets reduced by about one order of magnitude, with a similar situation to the SM annihilation background. The NC/CC VBF backgrounds get a two/one-order of magnitude reduction.

\begin{figure}[tb]
\centering
\includegraphics[width=0.45\textwidth]{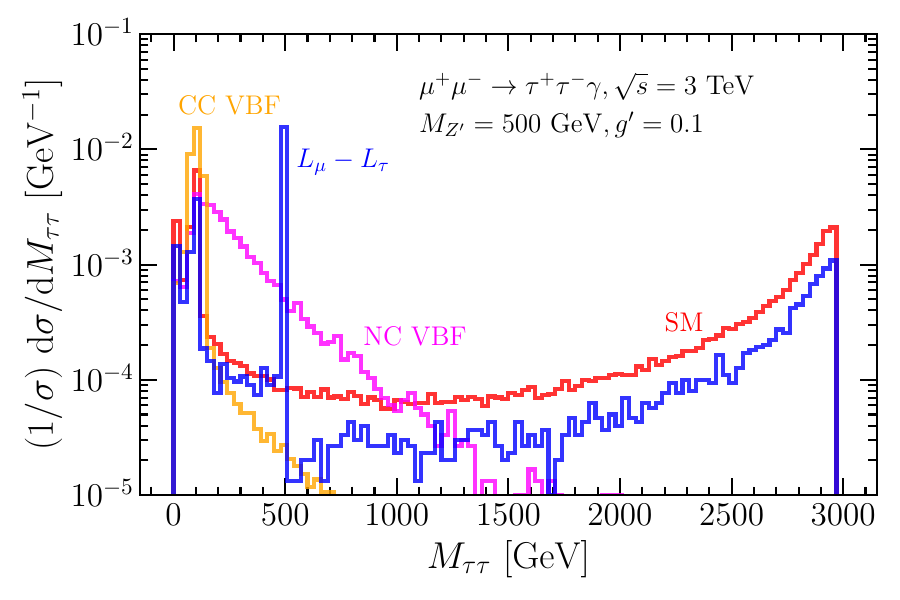}
\includegraphics[width=0.45\textwidth]{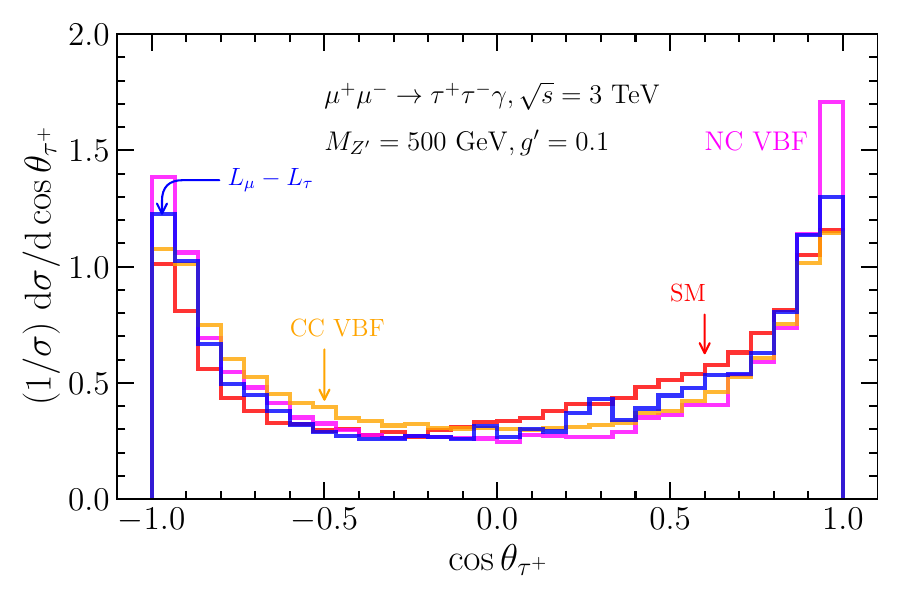}
\includegraphics[width=0.45\textwidth]{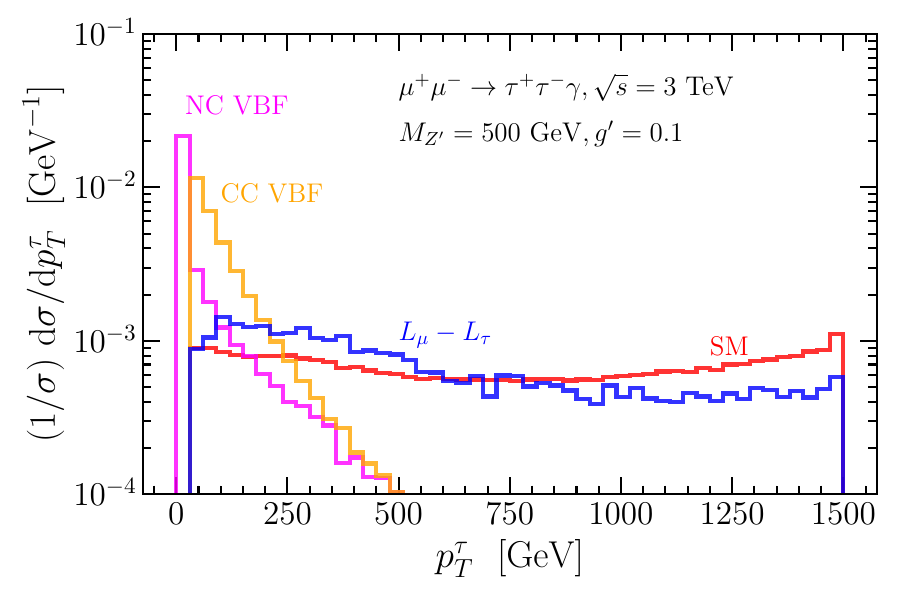}
\includegraphics[width=0.45\textwidth]{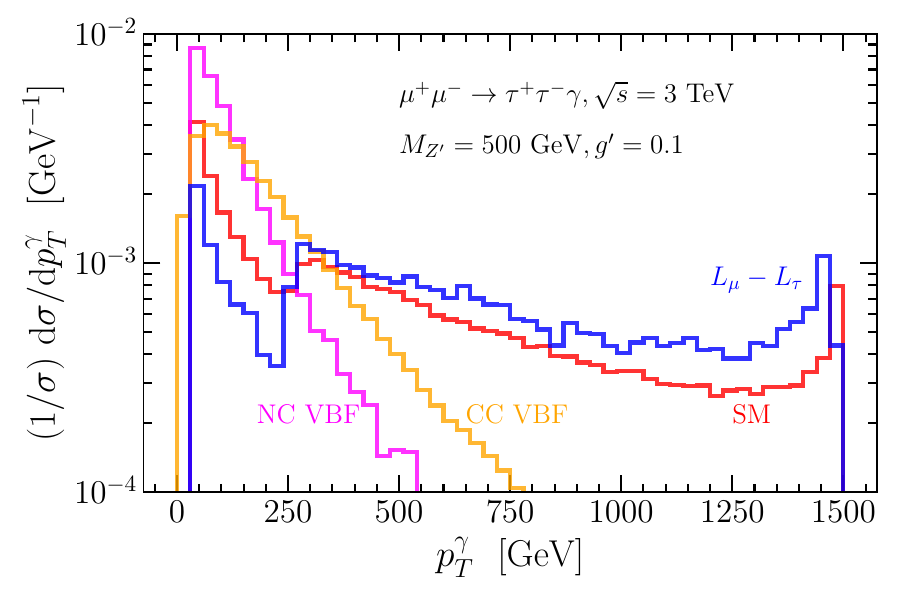}
\includegraphics[width=0.45\textwidth]{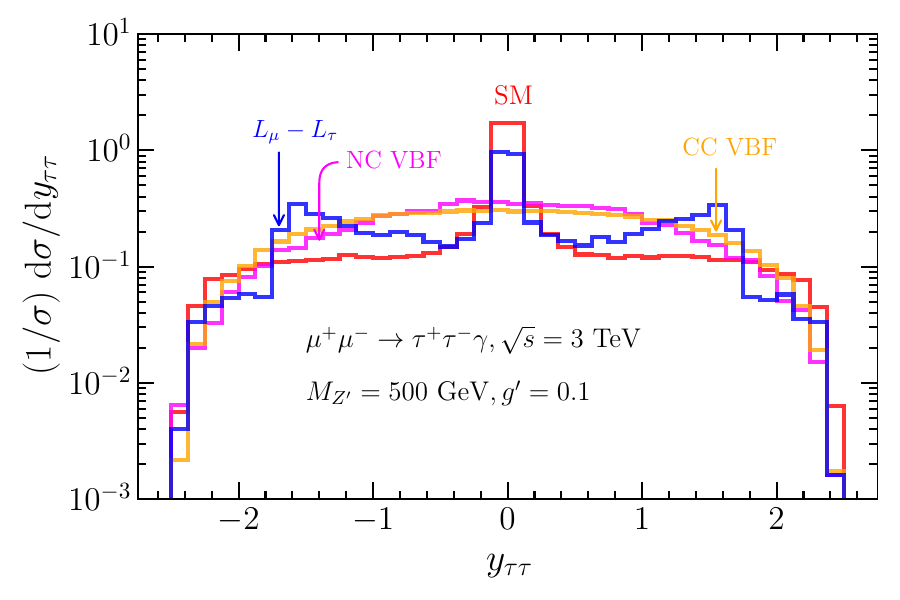}
    \caption{The distributions $\cos\theta_{\tau^+}$, $p_T^{\tau^+}$, $M_{\tau\tau}$  and $p_T^\gamma$ in the $\mu^+\mu^-\to\tau^+\tau^- \gamma$ scattering process at a 3 TeV muon collider with PSCs, Eq.~(\ref{eq:PSC}) and Eq.~(\ref{eq:PSC2}).
    }
\label{fig:dist_tatagama}
\end{figure}

In Fig.~\ref{fig:dist_tatagama}, we show the distributions of the lepton-pair invariant mass $M_{\tau\tau}$, rapidity $y_{\tau\tau}$, the cosine angle of the $\tau^+$, as well as the transverse momentum for $\tau^+$ and photon $\gamma$, with $M_{Z'}=500~\GeV$ for demonstration.
As before, we get resonance peak at $M_{\tau\tau}=M_{Z'}$, which motivates an optimization cut $|M_{\tau\tau}-M_{Z'}|<0.05M_{Z'}$, with the cut efficiency shown in Table \ref{tab:cutflow2}, and signal and background cross-sections shown in Fig.~\ref{fig:llA3TeV} (right). 
Meanwhile, we also see minor side peaks in the rapidity distributions around $|y(\eta)|\sim1.67$, which can be directly read from Eq.~(\ref{eq:yZp}) with the boundary condition $|\eta|=2.44$. But the spreading is much wider than the inclusive $Z'$ channel, due to the hard photon recoiling.

\begin{table}[tb]
\centering
\begin{tabular}{c|c|c|c||c|c|c}
\hline\hline
$U(1)_{L_\mu-L_\tau}$ & \multicolumn{3}{c||}{Signal} & \multicolumn{3}{c}{SM}\\
\hline
$M_{Z'}~[\TeV]$ & 0.5 & 2 & 5 & Annihilation & CC VBF & NC VBF\\
\hline
$\sigma$ [fb] & \multicolumn{5}{c}{$\mu^+\mu^-\to\tau^+\tau^-\gamma$} \\
\hline\hline
% & 8.16 & 14.4 & 4.17 & 4.04 & 5.66 & 17.27 \\
%$|\eta_\ell|<2.44$ & 7.50 & 13.7 & 3.92 & 3.80 & 4.70 & 8.94 \\
Eqs.~(\ref{eq:PSC}) and (\ref{eq:PSC2})  & 8.37& 14.6 & 5.26 & 4.93 &  2.29 & 1.14 \\    
%$M_{\ell\ell}>100~\GeV$ & 0.843 & 9.88 & 1.26 & 1.27 & 227 \\
$0.475<M_{\ell\ell}/\TeV<0.525$ & 3.95 &  &  & $2.16\cdot10^{-2}$ & $6.95\cdot10^{-3}$ & $4.85\cdot10^{-2}$ \\
$1.9<M_{\ell\ell}/\TeV<2.1$ & & 10.4 & & $6.95\cdot10^{-2}$ & $1.21\cdot10^{-2}$ & $2.70\cdot10^{-3}$ \\
$M_{\ell\ell}>0.95\sqrt{s}$ &  &  & 1.14 & 0.40& $2.50\cdot10^{-5}$ & $1.08\cdot10^{-3}$ \\
\hline\hline
\end{tabular}
\caption{The cut-flow table for $\tau^+\tau^-\gamma$ pair production at a $\sqrt{s}=3~\TeV$ muon collider. 
The gauge coupling is fixed at $g'=0.1$.
}
\label{tab:cutflow2}
\end{table}

%%%%%%%%%%%%%%%%%%%%%%%%%%%%%%%%%%%%%%%%
%\subsection{The mono-photon and four-lepton production}
%%%%%%%%%%%%%%%%%%%%%%%%%%%%%%%%%%%%%%%%

\begin{figure}[tb]
    \centering

\includegraphics{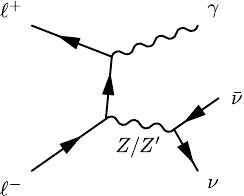}
   \includegraphics{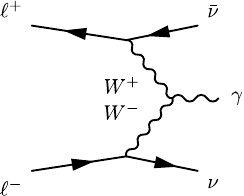}
    \includegraphics{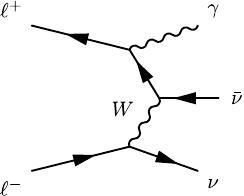}    
    \caption{Feynman diagrams for the mono-photon signal and background at a high-energy lepton collider.}
    \label{fig:FeynMonoV}
\end{figure}

\subsection{Mono-photon final state}
For the model under consideration, besides the charged-lepton channels, the $Z'$ can also decay into neutrinos, with the corresponding flavors.
Although this decay mode will result in missing momentum, the additional photon radiation, as shown in Fig.~\ref{fig:FeynMonoV}, 
will help to trigger the events and reconstruct the missing mass. 

We can take advantage of the ``recoil mass"~\cite{Han:2020pif} defined as 
\begin{equation}
M^2_{\rm recoil}=(p_1+p_2-p_\gamma)^2=s-2\sqrt{s}E_{\gamma}\, .
\end{equation}
For the on-shell $Z'$ production, the photon energy will be monochromatic and the recoil mass will lead to a mass peak at the $Z'$ resonance, in spite of the $Z'$ decay final states.

\begin{figure}[tb]
\centering
\includegraphics[width=0.49\textwidth]{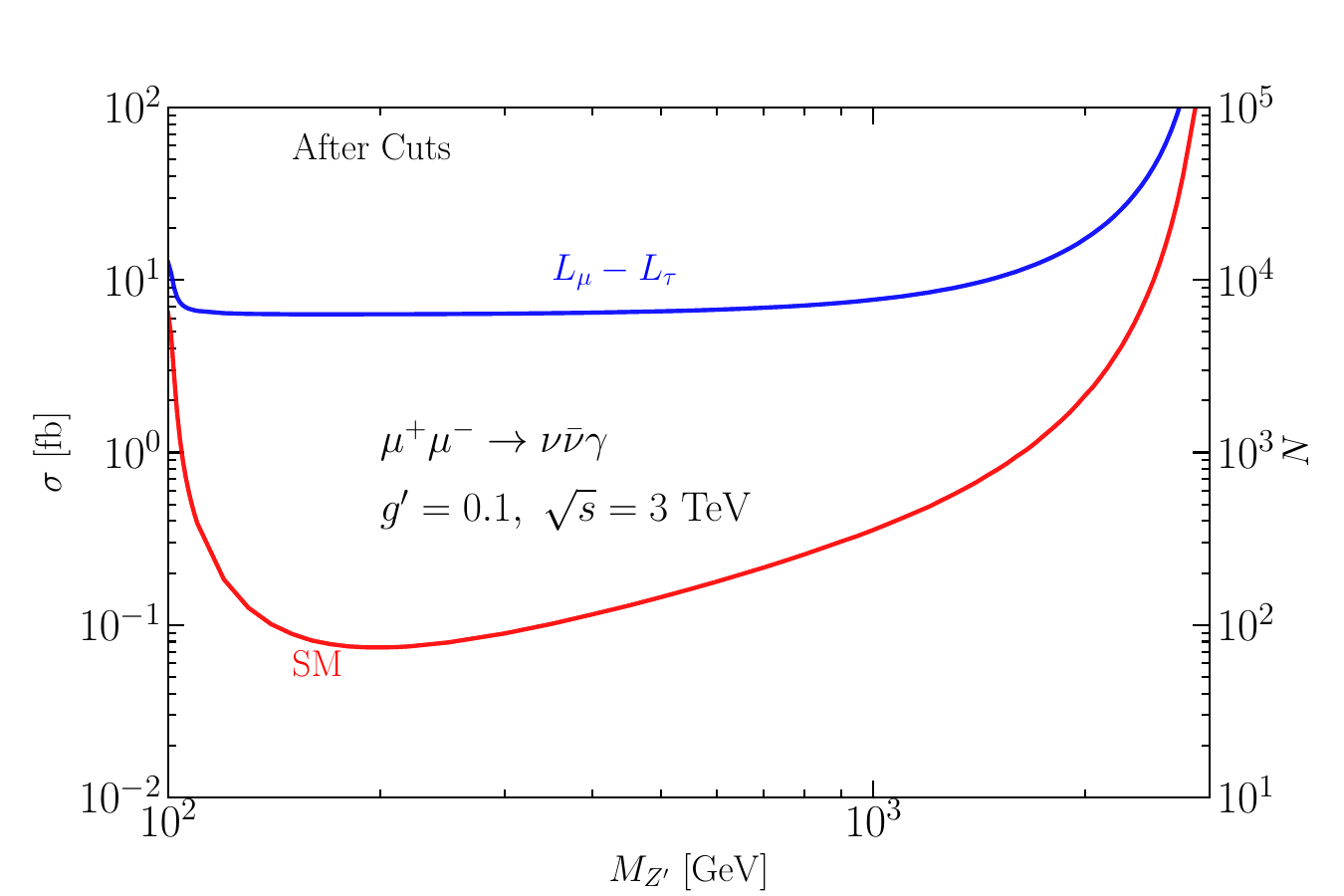}
\caption{The cross-sections for the mono-photon signal $\mu^+\mu^-\to\nu\bar{\nu}\gamma$ and SM background with cuts in Eq.~(\ref{eq:PSC2}) and $|M_{\rm recoil}-M_{Z'}|<10~\GeV$. 
}
\label{fig:xsmono}
\end{figure}

Similarly as before, the pre-selection cuts as Eq.~(\ref{eq:PSC2}) together with $M_{\rm recoil}=M_{\nu\bar{\nu}}>150~\GeV$ (to remove the on-shell $Z\to\nu\bar{\nu}$ background) are imposed. 
The photon energy is monochromatic near $E_\gamma = (s-M^2_{Z'})/2\sqrt s \approx 1460$ GeV for $M_{Z'}=500~\GeV$. Thus, the missing neutrinos lead to the recoil mass, as the reconstructed resonance peak around $M_{\rm recoil}=M_{Z'}$. 
Based on this observable, we can optimize the signal-background ratio with an additional cut $|M_{\rm recoil}-M_{Z'}|<10~\GeV$, with the efficiency demonstrated in Table \ref{tab:cutflow3}. The cross-sections for the signal and background are shown in Fig.~\ref{fig:xsmono}.

\begin{table}[tb]
\centering
\begin{tabular}{c|c|c|c||c}
\hline\hline
$U(1)_{L_\mu-L_\tau}$ & \multicolumn{3}{c||}{Signal} & \multicolumn{1}{c}{SM}\\
\hline
$M_{Z'}~[\TeV]$ & 0.5 & 2 & 5 & Annihilation\\
\hline
$\sigma$ [fb] & \multicolumn{4}{c}{$\mu^+\mu^-\to\nu\bar{\nu}\gamma$} \\
\hline\hline
$|\eta_\gamma|<2.44,p_T^\gamma>30~\GeV$ &$1.61\cdot10^{3}$  & $1.62\cdot10^{3}$ & $1.60\cdot10^{3}$ & $1.61\cdot10^{3}$    \\
$0.475<M_{\rm recoil}/\TeV<0.525$ & 7.00 &  &  &0.39    \\
$1.9<M_{\rm recoil}/\TeV<2.1$ & &38.6  & & 22.5    \\
$M_{\rm recoil}>0.95\sqrt{s}$ & &  & 837&  839   \\
\hline\hline
\end{tabular}
\caption{The cut-flow table for mono-photon production at a $\sqrt{s}=3~\TeV$ muon collider. 
The gauge coupling is fixed at $g'=0.1$. 
}
\label{tab:cutflow3}
\end{table}

\subsection{Four-lepton final states}
%%%%%%%%%%%%%%%%%%%%%%%%%%%%%%%%%%%%%%%%
\begin{figure}[tb]
    \centering
    \includegraphics{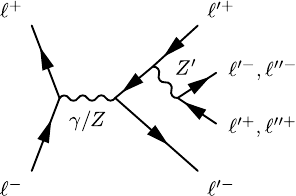}
    \caption{Feynman diagrams for four-lepton production via $Z'$ (which does not couple to the beam particles). %\BD{Please write $\gamma/Z$ for the first  mediator.}
    }
    \label{fig:Feyn4l}
\end{figure}

The discussions thus far rely on the fact that the $Z'$ couples to the initial $e^\pm$ or  $\mu^\pm$ beam particles at a collider. In the scenario that the $Z'$ does not have coupling to the initial beam particles, it can still be produced through the radiation of the final-state particle, as shown in Fig.~\ref{fig:Feyn4l}. 
Take the $U(1)_{L_\mu-L_\tau}$ model at an $e^+e^-$ collider as an example. The signal processes under consideration come from 
%$Z'\to \mu^+\mu^-, \tau^+\tau^-$ 
\begin{equation}
e^+e^-\to \mu^+\mu^- Z',\ \tau^+\tau^- Z' 
\to \mu^+\mu^-\mu^+\mu^-,\ \mu^+\mu^-\tau^+\tau^-,\ \tau^+\tau^-\tau^+\tau^- .
\end{equation}
In comparison with the scenario where $Z'$ directly couples to the initial beam particle, the sensitivity is expected to be much weaker, as a result of the smaller production cross-section and relatively higher SM background. 
As before, we take Eq.~(\ref{eq:PSC}) as pre-selection cuts and optimize with an invariant mass cut $|M_{\ell\ell}-M_{Z^\prime}|\leq 10~\GeV$ for electron/muon ($0.05M_{Z'}$ for $\tau$) pairs to present the sensitivity projections. We note that the leading SM background is from $e^+e^- \to ZZ(\gamma^*\gamma^*) \to \ell^+\ell^-\ \ell^+\ell^-$. Thus we can significantly enhance the signal sensitivity by removing this $Z/\gamma^*$ contribution with a cut $M(\ell'^+ \ell'^-)>150$ GeV for the lepton pair, which may be lowered and adjusted when very close to the threshold.

In fact, there are other contributions with $Z'$ radiating off the final state neutrinos: $e^+e^- \to Z^* \to \nu\bar \nu Z' \to \nu\bar \nu \tau^+\tau^-$. Similarly, we could further improve the sensitivity by including the additional decay channels $Z'\to \nu_\mu \bar \nu_\mu,\ \nu_\tau \bar \nu_\tau$. One could utilize the recoil mass variable from the accompanying charged leptons to reconstruct the $Z'$ resonant signal in this invisible decay mode. However, we expect such a larger background in the missing neutrino channels that we will not perform a comprehensive analysis here. 

%%%%%%%%%%%%%%%%%%%%%%%%%%%%%%%%%%
\subsection{Sensitivity summary}
%%%%%%%%%%%%%%%%%%%%%%%%%%%%%%%%%%
Now we will combine the sensitivities from all the channels discussed above and will summarize our results for the direct and indirect searches of the leptophilic $Z'$ model at high-energy lepton colliders. 

For the direct on-shell $Z'$ resonance production, 
%with $M_{Z'}<\sqrt{s}$, 
we use the statistical significance metric
\begin{equation}
\mathcal{S}=\frac{S}{\sqrt{S+B+\delta^2(S+B)^2}} \, ,
\end{equation}
and use $\mathcal{S}=2$ as the  $2\sigma$ sensitivity limit  (equivalent to $95\%$ CL) in  presenting our projections. Here, the systematic uncertainty is assumed to be  $\delta=0.1\%$~\cite{Han:2020pif}.
The $S$ and $B$ correspond to the signal and background events, respectively:
\begin{equation}
\begin{aligned}
&S=N^{\textrm{SM}+Z'}-N^{\textrm{SM}}=
\varepsilon\mathcal{L}(\sigma^{\textrm{SM}+Z'}-\sigma^{\textrm{SM}})\, , \\
&B=N^{\textrm{SM}}=\varepsilon\mathcal{L}\sigma^{\textrm{SM}} \, ,
\end{aligned}
\end{equation}
with $\varepsilon$ as the reconstruction efficiency. For illustration, we take a $\sqrt s=3$ TeV electron/muon collider with an integrated luminosity of $\mathcal{L}=1~\textrm{ab}^{-1}$. 
For electron and muon final states, the reconstruction efficiency can reach above $95\%$ at lepton colliders~\cite{ALEPH:2005ab}, and even close to 100\%~\cite{CLICdp:2018cto}.
In comparison, the tau identification efficiency can reach above 70\%~\cite{Li:2015sza} (and potentially 80\%~\cite{Yu:2020bxh}%\BD{Since this channel is not relevant here, why do we mention it?}
) at lepton colliders. In this work, we follow the treatment in Ref.~\cite{Huang:2021nkl} to apply $\varepsilon_{e,\mu}=100\%$ detection efficiency for electron and muon final-state events, while $\varepsilon_{\tau}=70\%$ for the final-state tau events, in addition to the larger invariant mass window cut as in Eq.~(\ref{eq:mzcut}).

For the indirect off-shell $Z$ production with $M_{Z'}>\sqrt{s}$, we take the cosine angle distribution with 20 even bins, as shown in Fig.~\ref{fig:costh}. We perform a bin-by-bin analysis with the $\chi^2$-sensitivity defined as
\begin{equation}
\chi^2=\sum_{i}\frac{S_i^2}{S_i+B_i+\delta^2(S_i+B_i)^2} \, ,
\end{equation}
where $S_i$ and $B_i$ respectively indicate the corresponding signal and background events in the $i^{\rm th}$ bin. We then use $\chi^2=4$ 
to obtain the $95\%$ CL sensitivity limit.

\begin{figure}
\centering
\includegraphics[width=0.49\textwidth]{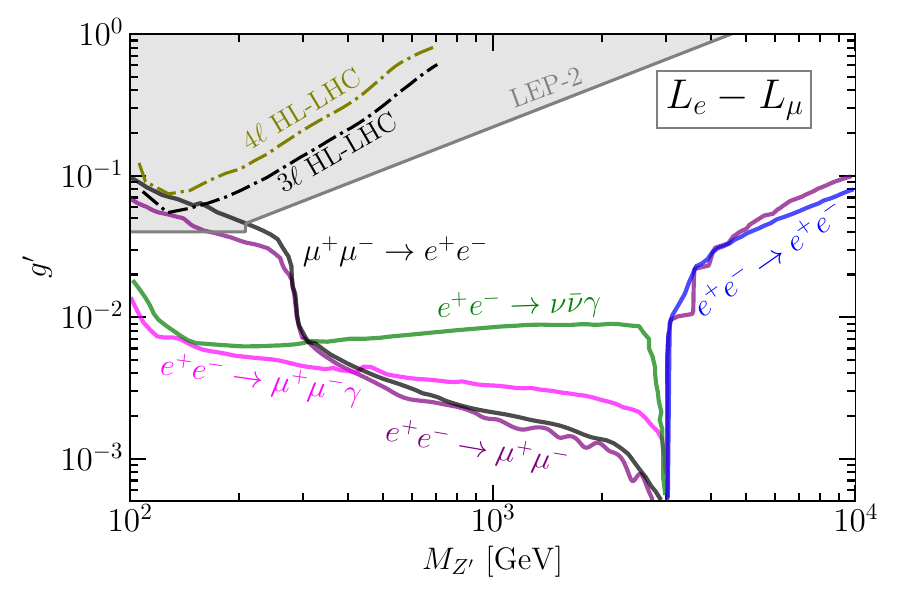}
\includegraphics[width=0.49\textwidth]{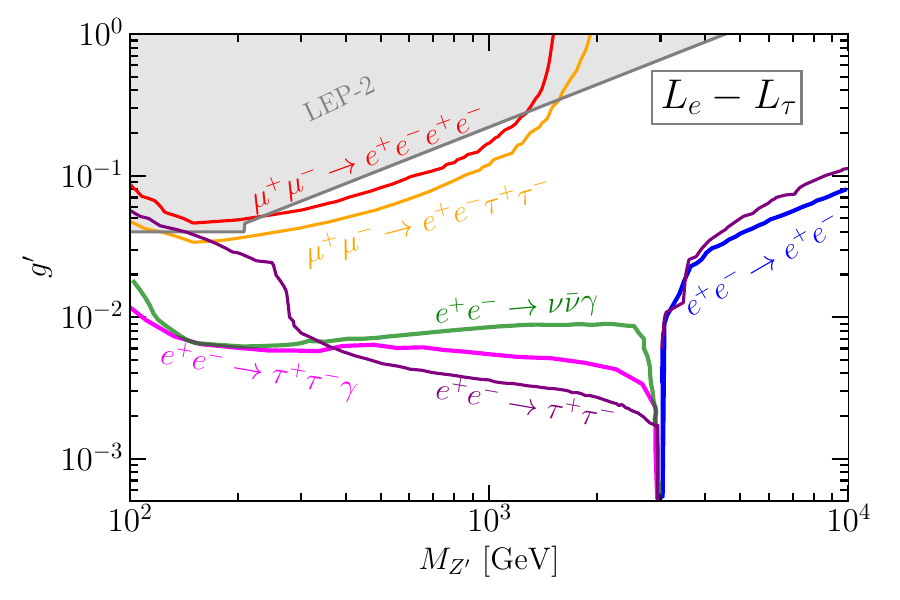}
\includegraphics[width=0.49\textwidth]{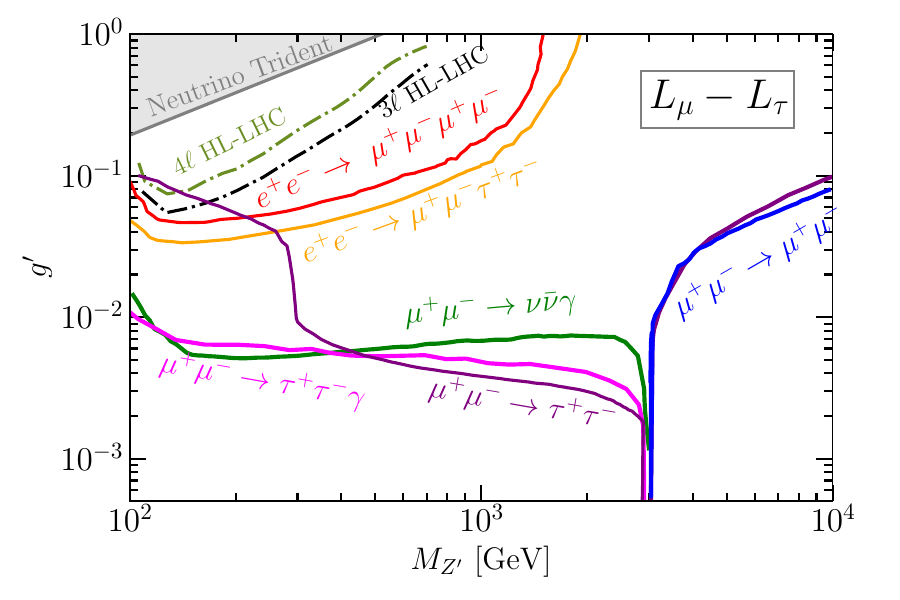}
\caption{The $95\%$ CL sensitivity for the $U(1)_{L_\alpha-L_\beta}$ models at $\sqrt{s} = 3~\TeV$ electron and muon colliders. The grey-shaded regions show the current exclusion limits [cf.~Fig.~\ref{fig:exist}]. We have also shown the projected sensitivity from $pp\to3\ell/4\ell$ at HL-LHC~\cite{delAguila:2014soa} by the black/green dot-dashed curves. 
}
\label{fig:sensi}
\end{figure}

The electron and muon collider sensitivities to the leptophilic $Z'$ models are summarized in Fig.~\ref{fig:sensi} for $\sqrt{s} = 3~\TeV$ with the optimal cuts discussed above.
The general features for the on-shell and off-shell probes of the $Z'$ parameter space are very similar, with the main difference arising from the tau final-state cut and sub-dominant differences from electron/muon beam mass effect and tau reconstruction efficiency with respect to the electron/muon ones. 

In each of the three cases of $U(1)_{L_\alpha-L_\beta}$, the strongest probe at large $Z'$ mass, {\it i.e.}, $M_{Z'}\gtrsim 300~\GeV$, comes from the lepton pair production through direct on-shell resonance decay with ISR, which reaches sensitivities down to $g'\sim10^{-3}$ when $M_{Z'}\sim \sqrt{s} = 3~\TeV$. The $U(1)_{L_e-L_\mu}$ model incorporates the best sensitivity among these three models, while the $U(1)_{L_e-L_\tau}$ sensitivity is slightly less due to the lower tau reconstruction efficiency, and the $U(1)_{L_\mu-L_\tau}$ sensitivity is further reduced because of the additional signal suppression from the ISR owing to the larger muon mass. In the low $Z'$ mass regime, the lepton-pair channel loses the constraining power because of two factors:  First, the VBF background, especially the NC one, takes over in the low mass region. Second, we have employed a rapidity optimization cut $|y_{\ell\ell}\pm y_{Z'}|<0.2$, where $y_{Z'}=\log(\sqrt{s}/M_{Z'})$, to improve the signal-to-background ratio, which loses effectiveness, when the $y_{Z'}$ goes beyond the detector acceptance.
As a consequence, the $\ell^+\ell^-\gamma$ associated production channel takes over in this regime, which was also observed in Ref.~\cite{Huang:2021nkl}.
In this regime, the mono-photon channel $\nu\bar{\nu}\gamma$ with the recoil mass can reach a similar sensitivity as shown in Fig.~\ref{fig:sensi}. In comparison between the electron and muon beams, the muon collider gets a slightly better sensitivity, due to the lower SM background induced by the larger muon mass in the photon radiation as Fig.~\ref{fig:FeynMonoV} shown. A similar situation happens to the photon associated production channel as well.

When $M_{Z'}$ goes above the collider energy $\sqrt{s}$, the resonance $Z'$ can be only produced off-shell. We have applied a bin-by-bin analysis based on the final-state angular distributions for both $s$- and $t$-channel processes, which provides an indirect probe to the $Z'$ coupling around $10^{-2}\sim10^{-1}$ up to $M_{Z'}<10~\TeV$, which improves upon the existing LEP contact interaction bound by roughly two orders of magnitude.
In general, we see that the $t$-channel Bhabha scattering gives a stronger probe than the $s$-channel annihilation, due to the modification of the FBA as shown in Fig.~\ref{fig:costh}. The $\tau$ final state, which can be only produced through the $s$ channel, gives a slightly worse sensitivity, as a consequence of the lower reconstruction efficiency than electrons and muons.

In the scenario that the gauge boson $Z'$ does not couple to the initial beam leptons, {\it e.g.}, for the $U(1)_{L_\mu-L_\tau}$ $Z'$ search at the electron collider, we focus on the four-lepton production channel, which can potentially produce the $Z'$ boson through the final-state radiation, {\it e.g.}, $e^+e^-\to \mu^+\mu^-(Z'\to\mu^+\mu^-/\tau^+\tau^-)$. A similar analysis based on the signal-to-background ratio has been performed, which shows a weaker sensitivity, around $5\times 10^{-2}\sim1$ for $100~\GeV<M_{Z'}\lesssim1~\TeV$, as shown in Fig.~\ref{fig:sensi}. It is driven by the smaller production rate, as well as a relatively larger SM background. 

To summarize, in comparison with the existing constraints, the future high-energy lepton colliders provide a great potential to probe extended regions of the leptophilic $Z'$ parameter space for $M_{Z'}>M_Z$.

\section{Gravitational wave signal}
\label{sec:GW}

As alluded to in Section~\ref{sec:intro}, the $U(1)$-extended gauge model, if classically conformal or scale-invariant,   guarantees the phase transition associated with its spontaneous breaking to be strongly first-order, and thus may lead to an observable GW signal~\cite{Jinno:2016knw, Chao:2017ilw, Brdar:2018num, Marzo:2018nov, Hasegawa:2019amx, Ellis:2020nnr, Huang:2022vkf, Dasgupta:2022isg, Chikkaballi:2023cce}. The GW signal is not only complementary to the searches at future lepton colliders discussed above, but can also probe an extended ($M_{Z'}, g')$ parameter space well beyond the reach of colliders. 
In this section, we explore the predictions for the GW signal in the conformal $U(1)_{L_\alpha-L_\beta}$ models, and show the complementarity with the collider constraints discussed above. For technical details of the formalism to compute the GW spectrum from strong first-order phase transition (SFOPT), see e.g.~Ref.~\cite{Athron:2023xlk}. 

%%%%%%%%%%%%%%%%%%%%%%%%%%%%%%%%%%
\subsection{Effective potential and thermal corrections}
\label{sec:potential}
%%%%%%%%%%%%%%%%%%%%%%%%%%%%%%%%%%    
Imposing the classically conformal invariance, the tree-level scalar potential at zero temperature is given as
\begin{equation}
    V_{\rm tree} = \lambda_H (H^\dag H)^2 + \lambda (\Phi^\dag \Phi)^2 - \lambda^\prime (\Phi^\dag \Phi)(H^\dag H) \, ,
    \label{eq:pot}
\end{equation}
where $H$ is the SM Higgs doublet, $\Phi=(\phi+iG)/\sqrt 2$ is an SM-singlet complex scalar field  which is responsible for the $U(1)$-symmetry breaking (see Table~\ref{tab:Model}), and we have assumed $\lambda_H, \lambda,\lambda'>0$. The quadratic mass terms (like $\mu_H^2 H^\dag H$) are absent in Eq.~\eqref{eq:pot} due to the  conformal invariance. In this case, the $U(1)$ symmetry breaking is achieved radiatively, {\it i.e.}, a non-zero VEV of $\Phi$, $\langle \Phi \rangle = v_\Phi/\sqrt 2$, arises purely from the renormalization group (RG) running of the quartic coupling $\lambda$, as in the original Coleman-Weinberg model~\cite{Coleman:1973jx}. 
This consequently gives mass to the $U(1)$ gauge boson,  $M_{Z^\prime} = 2g^\prime v_\Phi$, 
as well as the tree-level mass term for the SM Higgs boson through the quartic term $\lambda'$, i.e. $m_h^2=\lambda'v_\Phi^2=2\lambda_Hv^2$~\cite{Iso:2009ss, Iso:2009nw}. The negative sign in the last term of Eq.~\eqref{eq:pot} ensures that the induced squared mass for the Higgs doublet is negative, and the electroweak symmetry breaking is driven in the same way as in the SM. 

For $v_\Phi\gg v$, the symmetry breaking occurs first along the $\phi$ direction. Following the Gildener-Weinberg approach~\cite{Gildener:1976ih}, the zero-temperature effective potential for $\phi$ can be written as \cite{Sher:1988mj, Meissner:2008uw} 
\begin{equation}
    V_0(\phi, t) = \frac{1}{4}\lambda(t)[G(t)]^4\phi^4 \, ,
\end{equation}
where $t=\log(\phi/\mu)$, with $\mu$ being the renormalization scale, and 
\begin{equation}
    G(t) = \exp[-\int^t_0 \dd t' \gamma(t')]\  .
\end{equation}
The anomalous dimension in the Landau gauge\footnote{The issues concerning the gauge-dependence and impact on GW predictions have been addressed in Refs.~\cite{Patel:2011th, Wainwright:2011qy, Garny:2012cg, Chiang:2017zbz, Lofgren:2021ogg}.} is given by
\begin{equation}
\gamma(t) = \frac{-a_2}{32\pi^2}g^{\prime 2}(t) \ , 
\end{equation}
with $a_2 = 24$~\cite{Iso:2009ss, Hashimoto:2014ela}. The gauge coupling strength $\alpha_{g^\prime} = g^{\prime 2}/4\pi$ and quartic coupling strength $\alpha_\lambda = \lambda/4\pi$ obey the following RG equations
\begin{equation}
    2\pi\frac{\dd\alpha_{g^\prime}}{\dd t} = b\alpha^2_{g^\prime}(t) \ , \qquad 
    2\pi\frac{\dd\alpha_\lambda(t)}{\dd t} = a_1\alpha^2_\lambda(t) + 8\pi \alpha_\lambda(t)\gamma(t) + a_3 \alpha^2_{g^\prime}(t) \, , 
    \label{eq:RGE}
\end{equation}
where $b=16/3$, $a_1 = 10$, and $a_3 = 48$~\cite{Iso:2009ss}.\footnote{We have checked using {\tt SARAH}~\cite{Staub:2008uz} that only the value of $b$ is different for the $U(1)_{L_\alpha-L_\beta}$ case from the $U(1)_{B-L}$ case.}  Setting the renormalization scale $\mu$ to be the VEV $v_\Phi$ at the potential minimum $\phi=v_\Phi$ (i.e $\mu=v_\Phi$), or equivalently, $t=0$, the stationary condition 
\begin{equation}
    \left.\frac{\dd V}{\dd\phi}\right|_{\phi=v_\Phi}=\frac{e^{-t}}{v_\Phi}\left.\frac{\dd V}{\dd t}\right|_{t=0}=0 \, ,
\end{equation}
leads us to the relation
\begin{equation}
    a_1 \alpha^2_\lambda(0) + a_3 \alpha^2_{g^\prime}(0) + 8\pi \alpha_\lambda(0) = 0  . 
    \label{eq:constraint}
\end{equation}
\noindent
Thus $\alpha_\lambda(0)$ can be determined by $\alpha_{g^\prime}(0)$, and therefore, the scalar sector has only two free parameters, $v_\Phi$ and $\alpha_{g'}(0)$, which can be traded for the gauge boson mass $M_{Z'}$ and the gauge coupling $g'$ evaluated at $\mu=v_\Phi$.  One can then analytically solve the running of the couplings, and hence, the scalar potential~\cite{Iso:2009ss}: 
\begin{equation}
\label{eq_V0}
    V_0(\phi,t) = \frac{\pi \alpha_\lambda(t)}{\left(1 - \frac{b}{2\pi}\alpha_{g^\prime}(0)t\right)^{a_2/b}} \ \phi^4 . 
\end{equation}

As for the finite-temperature effects on the effective scalar potential, since the time evolution has two scales in it, $\phi$ and $T$ (with $T$ being the temperature of the Universe), we replace the renormalization scale parameter $t$ with $u = \log(\Lambda/v_\Phi)$, where $\Lambda =  {\rm max}(\phi,T)$ represents the largest energy scale in the system. The finite-temperature, one-loop effective potential is then given by 
\begin{equation}
    V_{\rm eff}(\phi,T) = V_0(\phi,u) + V_T(\phi,T) + V_{\rm daisy}(\phi, T) \, ,
    \label{eq:Veff}
\end{equation}
where $V_0$ is given by Eq.~\eqref{eq_V0}, and $V_T$ is the thermal contribution from bosonic one-loop~\cite{Jinno:2016knw}:
\begin{equation}
    V_T(\phi,T) = \frac{3T^4}{2\pi^2}J_B\left(\frac{M_{Z'}^2(\phi)}{T^2}\right) \, ,
    \end{equation}
where the bosonic thermal function is  
    \begin{equation}
    J_B(x) = \int^\infty_0\dd z \: z^2 \log \left[1-\exp(-\sqrt{z^2+x})\right] \, .
    \end{equation}
To improve the perturbative analysis beyond leading-order, we include in Eq.~\eqref{eq:Veff} the corrections due to the resummation of daisy diagrams~\cite{Arnold:1992rz}:  
    \begin{equation}
    V_{\rm daisy}(\phi,T) = \frac{T}{12\pi}\left[M^3_{Z'}(\phi)-M^3_{Z'}(\phi,T) \right] \, , 
\end{equation}
where the field- and temperature-dependent gauge boson masses are given by\footnote{Here, we have neglected the contribution from the $\lambda$-term to the thermal loop, since it is much smaller than the one from gauge interaction. } 
\begin{equation}
    M_{Z'}(\phi)  = 2g'\phi \, , \qquad 
    M_{Z'}(\phi,T)  = \sqrt{M_Z'^2(\phi)+\Pi_{Z'}^2(T)} \, ,
\end{equation}
where $\Pi_{Z'}(T)=2g'T$ is the thermal mass of the longitudinal component of the $Z'$ boson. 
%%%%%%%%%%%%%%%%%%%%%%%%%%%%%%%%%%
\subsection{Strong first-order phase transition}
\label{sec:phase}
%%%%%%%%%%%%%%%%%%%%%%%%%%%%%%%%%%
\begin{figure}[tb]
    \centering   \includegraphics[width=0.4\textwidth]{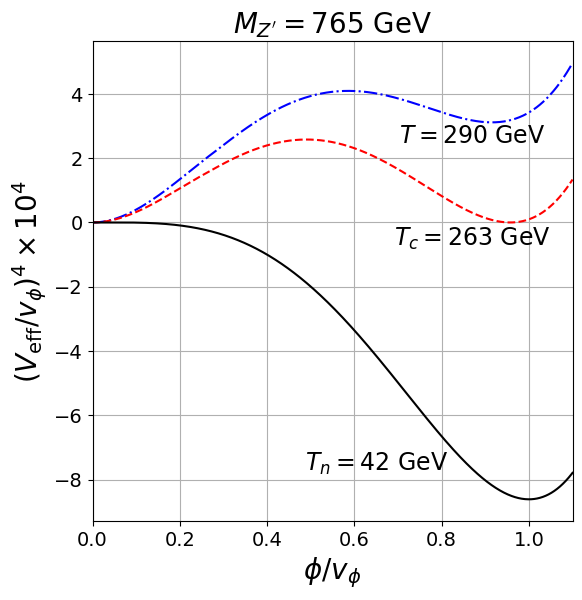}
    \caption{
    Effective potential at different temperatures $T>T_c$ (dot-dashed), $T=T_c$ (dashed) and $T=T_n$ (solid), for a fixed $M_{Z^\prime} = 765$ GeV. 
    }
    \label{fig:potential}
\end{figure}
To study the cosmological evolution of the effective potential~\eqref{eq:Veff}, it is useful to approximate it as~\cite{Jinno:2016knw} 
\begin{equation}
    V_{\rm eff}(\phi,u) \simeq \frac{1}{4}\lambda_{\rm eff}(u)\phi^4+\frac{1}{2}g'^2(u)T^2\phi^2 \, ,
\end{equation}
with $\lambda_{\rm eff}=4\pi\alpha_\lambda(u)/\left( 1-\frac{b}{2\pi}\alpha_{g^\prime}(0)u \right)^{a_2/b}$ [cf.~Eq.~\eqref{eq_V0}]. For $T \gg v_\Phi$, 
the effective potential has a unique minimum at $\phi = 0$. For $T \ll v_\Phi$, 
the self-coupling $\lambda_{\rm eff}$ around $\phi\lesssim T$ becomes negative, and 
therefore, 
 $\phi=0$ becomes a false vacuum. This takes place at the critical temperature 
 $T_c\approx (2/3) g' v_\Phi$. 
We show the evolution of the effective potential at a few representative temperatures versus the field strength normalized 
by $v_\Phi$ in Fig.~\ref{fig:potential} (where for illustration, we take $M_{Z^\prime} = 765$ GeV, which gives $T_c=263$ GeV).
For temperatures $T>T_c$, the 
true minimum is at $V_{\rm eff}=0$, as shown by the dot-dashed curve.
%in Fig.~\ref{fig:potential}. 
At $T=T_c$ (dashed curve), the two minima are 
degenerate. As the temperature of the Universe drops below $T_c$, the minimum at $\phi=0$ becomes the false vacuum $\phi_{\rm false}$ (solid curve). At the nucleation temperature $T_n$ (to be defined below), the field $\phi$ 
which is trapped around the false vacuum will tunnel to the 
true minimum, $\phi_{\rm true}$~\cite{Coleman:1977py}. This transition is  
first-order, provided the transition rate exceeds the expansion rate of the Universe. In 
this case, the transition triggers bubble nucleation, and subsequent GW production. 

%%%%%%%%%%%%%%%%%%%%%%%%%%%%%%%%%%%%%%%%%%%%%%%%%%
%\subsubsection{Nucleation rate}
%%%%%%%%%%%%%%%%%%%%%%%%%%%%%%%%%%%%%%%%%%%%%%%%%%
The nucleation rate per unit volume  is 
given by~\cite{Linde:1977mm,Linde:1981zj} 
\begin{equation}
    \Gamma(T) = [A(T)]^4\exp[-S(T)] \, ,
    \label{eq:bubble}
\end{equation}
where $A$ is a pre-factor of mass dimension one (see below), and $S$ is the Euclidean bounce action. At zero temperature, the configuration minimizing the action is $O(4)$-symmetric, and $S\equiv S_4$, where 
\begin{equation}
 S_4 = 2\pi^2 \int_0^\infty \dd r\: r^3 \left[\frac{1}{2}\left(\frac{\dd\phi}{\dd r}\right)^2+V_{\rm eff}(\phi,0)\right] \, ,
\label{eq:S4}   
\end{equation}
and can be estimated using the saddle-point approximation from the equation of motion:
\begin{equation}
    \frac{\dd^2 \phi}{\dd r^2} + \frac{3}{r} \frac{\dd\phi}{\dd r} - \frac{\partial V_{\rm eff}}{\partial \phi}
= 0 \, ,
\end{equation}
with the boundary conditions
\begin{equation}
    \frac{\dd\phi}{\dd r}(r=0)=0 \, , \qquad \phi(r=\infty)=0 \, .
    \label{eq:boundary}
\end{equation}
In the above equation, the first condition is for the solution to be regular at the center of the bubble, and the second one is to describe the initial false vacuum background far from the bubble. The bubble nucleation rate is eventually well approximated at low $T$ by $\Gamma\equiv \Gamma_4$, where [cf.~Eq.~\eqref{eq:bubble}]
\begin{equation}
    \Gamma_4 \simeq \frac{1}{R_c^4}\left(\frac{S_4}{2\pi}\right)^2\exp(-S_4) \, ,
\end{equation}
with $R_c\sim 1/T$ being the bubble radius in the low $T$ limit.  

At finite temperature, the field becomes periodic in the time coordinate (or in $1/T$). The configuration minimizing the action in this case is $O(3)$-symmetric. Moreover, at sufficiently high temperatures, the minimum action configuration becomes constant in the time direction and $S\equiv S_3(T)/T$, where 
\begin{equation}
S_3(T) = 4\pi \int_0^\infty \dd r \: r^2 \left[\frac{1}{2}\left(\frac{\dd\phi}{\dd r}\right)^2+\Delta V_{\rm eff}(\phi,T)\right] \, ,
\label{eq_S3}
\end{equation}
where $\Delta V_{\rm eff}(\phi,T) \equiv V_{\rm eff}(\phi,T) - V_{\rm eff}(\phi_{\rm false},T)$. Eq.~\eqref{eq_S3} represents bubble formation through classical field excitation over the barrier, with the corresponding equation of motion given by 
\begin{equation}
\frac{\dd^2 \phi}{\dd r^2} + \frac{2}{r} \frac{\dd\phi}{\dd r} - \frac{\partial V_{\rm eff}}{\partial \phi}
= 0 \, ,
\label{eq:saddle}
\end{equation}
with the same boundary conditions as in Eq.~\eqref{eq:boundary}. The solution to Eq.~\eqref{eq:saddle} extremizes the action~\eqref{eq_S3} that gives the exponential suppression of the false vacuum decay rate~\cite{Coleman:1977th}. From Eq.~\eqref{eq:bubble}, the nucleation rate $\Gamma\equiv \Gamma_3$ can be calculated as 
\begin{equation}
\Gamma_3 \simeq T^4\left(\frac{S_3(T)}{2\pi T}\right)^{3/2} \exp\left[-\frac{S_3(T)}{T}\right] \, .
\label{eq_Gamma1}
\end{equation}
In practice, the exact solution with a non-trivial periodic bounce in the time coordinate, which corresponds to quantum tunneling at finite temperature, is difficult to evaluate. Following Ref.~\cite{Espinosa:2018hue}, we have taken the minimum of the two actions $S_3$ and $S_4$ in our numerical calculation of the bubble nucleation rate, {\it i.e.}, $\Gamma\approx {\rm max}(\Gamma_3,\Gamma_4)$. For a discussion of related theoretical uncertainties, see Ref.~\cite{Croon:2020cgk}.

The nucleation temperature is defined as the inverse time of creation of one bubble per Hubble radius, {\it i.e.},
\begin{equation}
    \left.\frac{\Gamma(T)}{H(T)^4}\right|_{T=T_n} =1 \, ,   \label{eq:Tn} 
\end{equation}
where the Hubble expansion rate at temperature $T$ is $H(T)\simeq 1.66\sqrt{g_*} T^2/M_{\rm Pl}$, with $g_*\simeq 110$ being the relativistic degrees of freedom at high temperatures,\footnote{In our numerical analysis, we take into account the temperature-dependence of $g_*$~\cite{Saikawa:2018rcs}.} and $M_{\rm Pl}\simeq 2.4\times 10^{18}$ GeV being the reduced Planck mass. 
%In practice, $T_n$ is often computed using the Hubble rate in the false vacuum for strong super-cooled phase transitions~\cite{Hambye:2018qjv, Baldes:2023rqv}. 

\begin{figure}[t!]
    \centering
    \begin{tabular}{lr} 
    \includegraphics[width=0.4\textwidth]{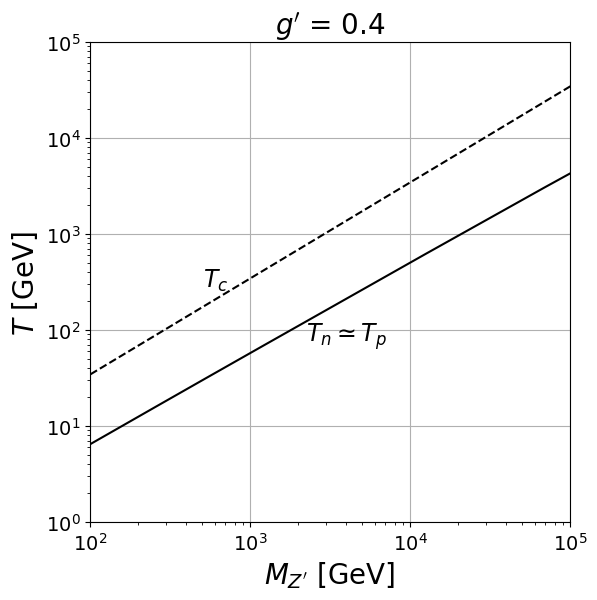} &    \includegraphics[width=0.4\textwidth]{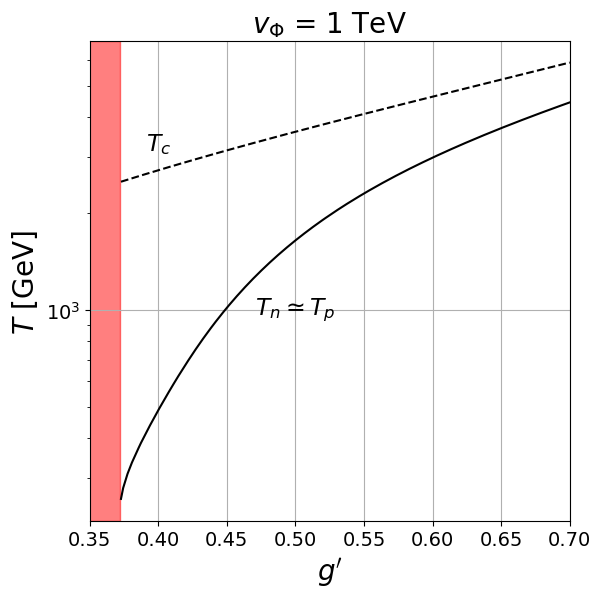} \\
    \includegraphics[width=0.38\textwidth]{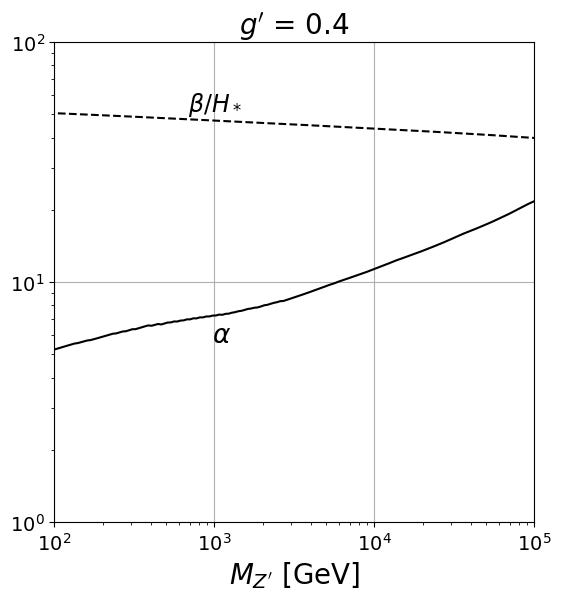} & \includegraphics[width=0.38\textwidth]{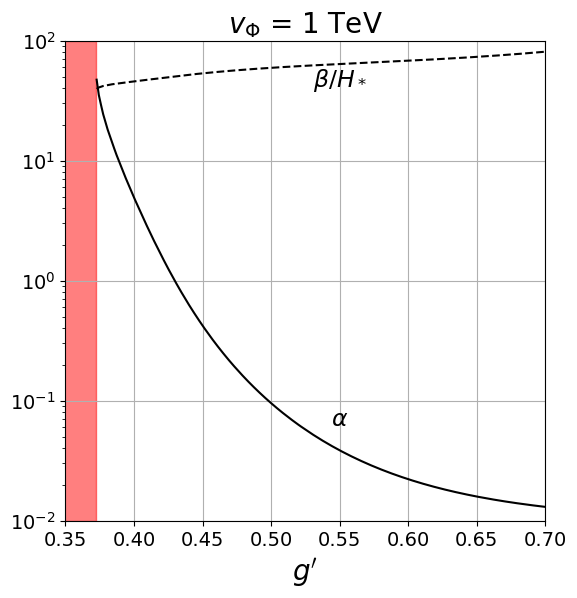}
    \end{tabular}
      \caption{Variation of the temperatures $T_c$, $T_n$ and $T_p$ (upper panels) and the phase transition parameters $\alpha$ and $\beta/H_*$ (lower panels) with respect to the model parameters $M_{Z'}$ (left) and $g'$ (right), keeping the other ones fixed at the values shown in the plots. The red region on the right panels corresponds to the lower limit on the gauge coupling below which the rate of phase transition is not fast enough with respect to the Hubble rate $H$ to achieve bubble nucleation.  }
    \label{fig:GWparam}
\end{figure}

%%%%%%%%%%%%%%%%%%%%%%%%%%%%%%%%%%%%%%%%%%%%%%%%%%
%\subsubsection{Vacuum Transition Probability}
%%%%%%%%%%%%%%%%%%%%%%%%%%%%%%%%%%%%%%%%%%%%%%%%%%
The statistical analysis of the subsequent evolution of bubbles in the early Universe is crucial for SFOPT~\cite{Ellis:2018mja}.  The probability
for a given point to remain in the false vacuum is given by $P(T) = \exp[-I(T)]$~\cite{Guth:1979bh, Guth:1981uk, Guth:1982pn,Enqvist:1991xw}, where $I(T)$ is the expected volume of true vacuum bubbles per comoving volume:
\begin{equation}
    I(T) = \frac{4\pi}{3}\int^{T_c}_{T}\frac{\dd T^\prime}{T^{\prime 4}}\frac{\Gamma(T^\prime)}{H(T^\prime)}\left(\int^{T^\prime}_{T}\frac{\dd\tilde{T}}{H(\tilde{T})}\right)^3.
\end{equation}
The change in the physical volume of the false vacuum, ${\cal V}_{\rm false}=a^3({\rm t})P(T)$ (with $a({\rm t})$ being the scale factor and ${\rm t}$ being the time), normalized to the Hubble rate, is given by~\cite{Turner:1992tz} 
\begin{equation}
    \frac{1}{{\cal V}_{\rm false}}\frac{\dd{\cal V}_{\rm false}}{\dd {\rm t}} = H(T)\left(3+T\frac{\dd I(T)}{\dd T}\right) \, .
\end{equation}
We define the percolation temperature $T_p$ as satisfying the condition $I(T_p)=0.34$~\cite{Ellis:2018mja}, while ensuring that the volume of the  false vacuum is decreasing,~ {\it i.e.},~$\dd\log{\cal V}_{\rm false}/\dd {\rm t}<0$, so that percolation is possible despite the exponential expansion of the false vacuum. Numerically, we find that the percolation temperature is only slightly smaller than the nucleation temperature, which in turn is smaller than the critical temperature. Moreover, as expected, both $T_c$ and $T_n\simeq T_p$ grow monotonically, as either of the model parameters, $M_{Z'}$ or $g'$, increases. They are depicted in Fig.~\ref{fig:GWparam}.

The strength of the phase transition is characterized by two quantities $\alpha$ and $\beta$ defined as follows: $\alpha=\epsilon_*/\rho_{\rm rad}$ is the ratio of the vacuum energy density $\epsilon_*$ released in the transition to the radiation energy density $\rho_{\rm rad}=\pi^2 g_* T_*^4/30$, both evaluated at $T=T_*$ (where $T_*$ is either $T_n$ or $T_p$).\footnote{Calculating $\alpha$ at $T_p$ is better, as it accounts for the entropy dilution  between $T_n$ and $T_p$, although this distinction becomes irrelevant for the GW signal when $\alpha\gg 1$. } The vacuum energy density is nothing but the free energy difference between the true and false vacua~\cite{Espinosa:2008kw}, thus yielding   
\begin{equation}
    \alpha =  \frac{1}{\rho_{\rm rad}}\left.\left(-1+T\frac{\dd}{\dd T}\right)\Delta V_{\min}\right|_{T=T_*} \, ,
    \label{eq:alpha}
\end{equation}
where $\Delta V_{\rm min}$ is the temperature-dependent minimum of the effective potential $\Delta V_{\rm eff}$ defined below Eq.~\eqref{eq_S3}. 

The second important parameter is $\beta/H_*$, where $\beta$ is the (approximate) inverse timescale of the phase transition and $H_*$ is the Hubble rate at $T_*$:
 \begin{equation}
     \frac{\beta}{H_*} = \left. -\frac{T}{\Gamma} \frac{\dd\Gamma}{\dd T}\right|_{T=T_*} \, .
     \label{eq:beta}
 \end{equation}
 For strong transitions, $\beta$ is related to the average bubble radius $R_*$: $\beta=(8\pi)^{1/3}/R_*$~\cite{Ellis:2020nnr},\footnote{When identifying $\beta/H_*$ with the average bubble radius, it is numerically found that taking $T_*=T_n$ gives a more accurate result~\cite{Baldes:2023rqv}.} where $R_*$ defines the characteristic length scale of transition and is given by~\cite{Turner:1992tz, Enqvist:1991xw}
 \begin{equation}
     R_* = \left[T_*\int_{T_*}^{T_c}\frac{\dd T}{T^2}\frac{\Gamma(T)}{H(T)}\exp\{-I(T)\}\right]^{-1/3} \, .
 \end{equation}
 The variation of $\alpha$ and $\beta/H_*$ with the model parameters $g'$ and $M_{Z'}$ is shown in Fig.~\ref{fig:GWparam}. We find that $\beta/H*$ decreases (increases) with increasing $M_{Z'}$ ($g'$), whereas $\alpha$ has the opposite behavior. Moreover, the change in $\alpha$ is more rapid than that in $\beta/H_*$. We also find that the gauge coupling cannot be decreased arbitrarily, because below a certain value (as shown by the red shaded region on the right panels of Fig.~\ref{fig:GWparam}), the rate of transition is not fast enough with respect to the Hubble rate $H$ to achieve bubble nucleation. 
%%%%%%%%%%%%%%%%%%%%%%
\subsection{Gravitational wave spectrum}
\label{sec:GWSFOPT}
%%%%%%%%%%%%%%%%%%%%%%
The amplitude of the GW signal as a function of the frequency $f$ is usually defined as 
\begin{equation}
h^2\Omega_{\rm GW}(f)\equiv \frac{h^2}{\rho_c}\frac{\dd\rho_{\rm GW}}{\dd\log f} \, ,
\end{equation}
where $h\sim 0.7$ is the dimensionless Hubble parameter (defined in terms of today's value of $H$, $H_0=100 h~{\rm km/s/Mpc}$), $\rho_{\rm GW}$ is the energy density released in the form of GWs, and $\rho_c=3H_0^2M_{\rm Pl}^2\simeq 1.05\times 10^{-5}h^2~{\rm GeV}/{\rm cm}^3$ is the critical density of the Universe. The reason for multiplying $\Omega_{\rm GW}$ by $h^2$ is to make sure that the GW amplitude is not affected by the experimental uncertainty~\cite{DiValentino:2021izs} in the Hubble parameter $H_0$. 

There are three different mechanisms for producing GWs in an SFOPT from the expanding and colliding scalar-field bubbles,  as well as from their interaction with the thermal plasma. These are: (i) collisions of expanding bubble walls~\cite{Kosowsky:1991ua,Kosowsky:1992rz,Kosowsky:1992vn,Kamionkowski:1993fg,Caprini:2007xq,Huber:2008hg,Weir:2016tov, Jinno:2017fby, Jinno:2019bxw,Lewicki:2020jiv,Megevand:2021juo}, compressional modes (or sound waves) in the bulk plasma~\cite{Hindmarsh:2013xza,Giblin:2013kea,Giblin:2014qia,Hindmarsh:2015qta, Hindmarsh:2016lnk,Hindmarsh:2017gnf,Hindmarsh:2019phv}, and (iii) vortical motion (or magnetohydrodynamic turbulence) in the bulk plasma~\cite{Caprini:2006jb,Kahniashvili:2008pf,Kahniashvili:2008pe,Kahniashvili:2009mf,Caprini:2009yp,Kisslinger:2015hua, RoperPol:2019wvy}. The total GW signal can be approximated as a linear superposition of the signals generated from these three individual sources, denoted respectively by $\Omega_b$ (bubble wall), $\Omega_s$ (sound wave), and $\Omega_t$ (turbulence):
\begin{equation}
    h^2\Omega_{\rm GW}(f) \simeq h^2\Omega_b(f) + h^2\Omega_s (f) + h^2\Omega_t(f) \, .
    \label{eq:tot_gw}
\end{equation}
The three contributions can be parameterized in a model-independent way in terms of a set of characteristic SFOPT parameters, namely,  $\alpha$ [cf.~Eq.~\eqref{eq:alpha}], $\beta/H_*$ [cf.~Eq.~\eqref{eq:beta}], $T_*$,\footnote{We will use $T_*=T_n$, the nucleation temperature defined in Eq.~\eqref{eq:Tn}. For subtleties, see Ref.~\cite{Athron:2022mmm}.} bubble-wall velocity $v_w$, and the three efficiency factors $\kappa_b$, $\kappa_s$,  $\kappa_t$ that characterize the fractions of the released vacuum energy that are converted into the energy of scalar field gradients, sound waves and turbulence,  respectively. The bubble-wall velocity in the plasma rest-frame is given by~\cite{Lewicki:2021pgr} 
\begin{equation}
    v_w = \left\{\begin{array}{ll}\sqrt{\frac{\Delta V_{\rm min}}{\alpha\rho_{\rm rad}}} & {\rm for}~\frac{\Delta V_{\rm min}}{\alpha\rho_{\rm rad}}<v_J \\
    1 & {\rm for}~\frac{\Delta V_{\rm min}}{\alpha\rho_{\rm rad}}\geq v_J \end{array}\right. \, ,
\end{equation}
where $v_J=(1+\sqrt{3\alpha^2+2\alpha})/\sqrt 3(1+\alpha)$ is the Jouguet velocity~\cite{Kamionkowski:1993fg, Steinhardt:1981ct,Espinosa:2010hh}. As for the efficiency factors, it is customary to express $\kappa_s$ and $\kappa_t$ in terms of another efficiency factor $\kappa_{\rm kin}$ that characterizes the energy fraction converted into bulk kinetic energy and an additional parameter $\varepsilon$, {\it i.e.}~\cite{Ellis:2019oqb} 
\begin{equation}
    \kappa_s = \kappa_{\rm kin} \, , \qquad \kappa_t = \varepsilon \kappa_{\rm kin} \, .
\end{equation}
While the precise numerical value of $\varepsilon$ is still under debate, following Refs.~\cite{Caprini:2015zlo,Hindmarsh:2015qta}, we will use $\varepsilon=1$. 
The efficiency factor $\kappa_b$ is taken from Ref.~\cite{Kamionkowski:1993fg} and $\kappa_{\rm kin}$ is taken from Ref.~\cite{Espinosa:2010hh}, both of which were calculated in the so-called Jouguet detonation limit:
\begin{equation}
    \begin{aligned}
        \kappa_b &= \frac{1}{(1 + 0.715 \alpha)}\left(0.715 \alpha + \frac{4}{27} \sqrt{\frac{3\alpha}{2}}\right) \, ,  \\
        \kappa_{\rm kin} &= \frac{\sqrt{\alpha}}{(0.135 + \sqrt{0.98 + \alpha})} \, .
    \end{aligned}
\end{equation}
Each of the three contributions in Eq.~\eqref{eq:tot_gw} is related to the SFOPT parameters discussed above, as follows~\cite{Schmitz:2020syl}:
\begin{equation}
    h^2 \Omega_i(f) = h^2\Omega^{\rm peak}_i\left(\alpha,\frac{\beta}{H_*},T_*,v_w,\kappa_i\right)\mathcal{S}_i(f,f_i) \, , 
    \label{eq:Omega}
\end{equation}
where $i\in\{b,s,t\}$, the peak amplitudes are given as~\cite{Caprini:2015zlo} 
\begin{equation}
\begin{aligned}
    h^2\Omega^{\rm peak}_b &\simeq 1.67\times 10^{-5} \left(\frac{v_w}{\beta/H_*}\right)^2\left(\frac{100}{g_*(T_*)}\right)^{1/3}\left(\frac{\kappa_b \alpha}{1+\alpha}\right)^2\left(\frac{0.11v_w}{0.42 + v^2_w}\right), \\
    h^2\Omega^{\rm peak}_s &\simeq 2.65\times 10^{-6} \left(\frac{v_w}{\beta/H_*}\right)\left(\frac{100}{g_*(T_*)}\right)^{1/3}\left(\frac{\kappa_s \alpha}{1+\alpha}\right)^2,  \\
    h^2\Omega^{\rm peak}_t &\simeq 3.35\times 10^{-4} \left(\frac{v_w}{\beta/H_*}\right)\left(\frac{100}{g_*(T_*)}\right)^{1/3}\left(\frac{\kappa_t \alpha}{1+\alpha}\right)^{3/2},
    \label{eq:Opeak}
\end{aligned}
\end{equation}
and the spectral shape functions are given as~\cite{Caprini:2015zlo} 
\begin{equation}
\begin{aligned}
    \mathcal{S}_b(f,f_b) &= \left(\frac{f}{f_b}\right)^{2.8}\left[\frac{3.8}{1+2.8(f/f_b)^{3.8}}\right],  \\
    \mathcal{S}_s(f,f_s) &= \left(\frac{f}{f_s}\right)^3\left[\frac{7}{4+3(f/f_s)^2}\right]^{7/2},  \\
    \mathcal{S}_t(f,f_t,h_*) &= \left(\frac{f}{f_t}\right)^3\left[\frac{1}{1+(f/f_t)}\right]^{11/3}\left(\frac{1}{1+8\pi f/h_*}\right)\, .
\end{aligned}
\end{equation}
Note that ${\cal S}_b$ and ${\cal S}_s$ are normalized to unity at their respective peak frequencies $f_b$ and $f_s$, whereas ${\cal S}_t$ at $f_t$ depends on the Hubble frequency 
\begin{equation}
    h_* = \frac{a_*}{a_0}H_*=1.6\times 10^{-2}~\textrm{mHz} \left(\frac{g_*(T_*)}{100}\right)^{1/6}\left(\frac{T_*}{100~\GeV}\right).
\end{equation}
And finally, the peak frequencies are given as
\begin{equation}
\begin{aligned}
    f_b &= 1.6\times10^{-2}~\textrm{mHz} \left(\frac{g_*(T_*)}{100}\right)^{1/6}\left(\frac{T_*}{100~\GeV}\right)\left(\frac{\beta/H_*}{v_w}\right)\left(\frac{0.62 v_w}{1.8 - 0.1 v_w + v^2_w}\right),  \\
    f_s &= 1.9\times10^{-2}~\textrm{mHz} \left(\frac{g_*(T_*)}{100}\right)^{1/6}\left(\frac{T_*}{100~\GeV}\right)\left(\frac{\beta/H_*}{v_w}\right),  \\
    f_t &= 2.7\times10^{-2}~\textrm{mHz} \left(\frac{g_*(T_*)}{100}\right)^{1/6}\left(\frac{T_*}{100~\GeV}\right)\left(\frac{\beta/H_*}{v_w}\right). \label{eq:fpeak}
\end{aligned}
\end{equation}
%%%%%%%%%%%%%%%%%%%%%%%%%%%%%
\begin{figure}[t!]
    \centering
\includegraphics[width=0.49\textwidth]{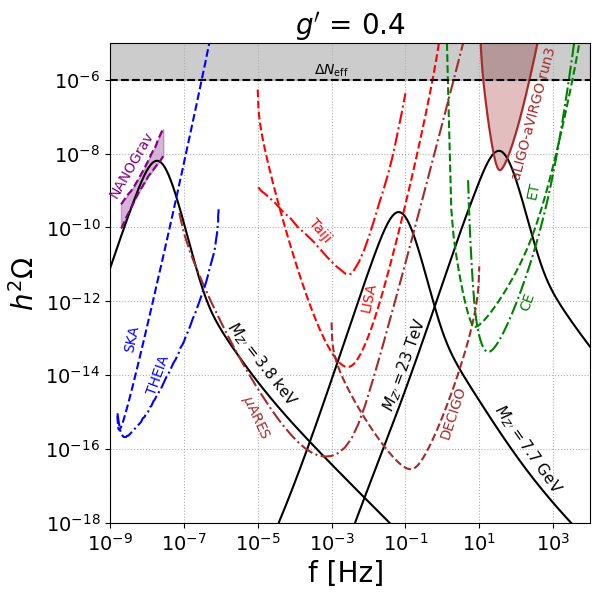}    \includegraphics[width=0.49\textwidth]{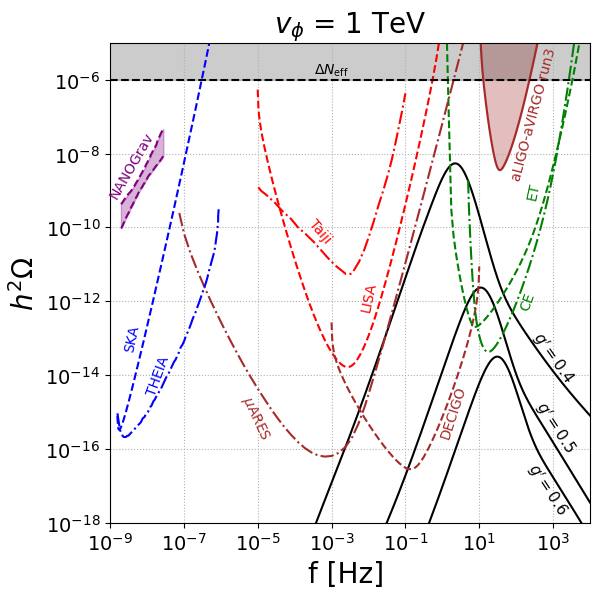}
    \caption{
    The stochastic GW amplitude in the $U(1)_{L_\alpha-L_\beta}$ models for different values of the $Z'$ mass $M_{Z'}$ (left panel) and gauge coupling $g'$ (right panel), while keeping the other parameter fixed at the value shown in the plot.  The current constraints from aLIGO-aVIRGO,  $\Delta N_{\rm eff}$, as well as the future sensitivities from planned GW experiments are shown for comparison. The recent NANOGrav observations in the nHz regime are also shown.  
    }
    \label{fig:GWplot}
\end{figure}
%%%%%%%%%%%%%%%%%%%%%%%%%%%%%

Since there are only two free parameters in our model setup, namely, $M_{Z'}$ (or $v_\Phi$) and $g'$, we show in Fig.~\ref{fig:GWplot} how the total GW amplitude as a function of the GW frequency varies with respect to these two model parameters.  In the left panel, we fix $g'=0.4$ and show the GW spectra for different values of $M_{Z'}$ (or equivalently, the VEV $v_\Phi$). It is clear that the whole spectrum shifts to higher frequencies with increasing $M_{Z'}$, which is due to the correlation of the symmetry-breaking scale with the nucleation temperature, which in turn moves the peak frequency [cf.~Eq.~\eqref{eq:fpeak}]. The peak amplitude increases with $M_{Z'}$, which is mostly due to its correlation with $\alpha$, and to a lesser extent, with $\beta/H_*$ [cf.~Eq.~\eqref{eq:Opeak} and Fig.~\ref{fig:GWparam}]. For very small $M_{Z'}$ values, the peak amplitude again starts to increase because of the smaller $g_*$. For comparison, we also show the current $95\%$ CL constraint on the stochastic GW amplitude from aLIGO-aVIRGO third run~\cite{KAGRA:2021kbb} (red shaded region on the upper right corner), as well as the $95\%$ CL  constraint on $\Delta N_{\rm eff}<0.18$ from a joint BBN+CMB analysis~\cite{Yeh:2022heq}, which translates into an upper bound on $h^2\Omega_{\rm GW}\leq 5.6\times 10^{-6}\Delta N_{\rm eff}$~\cite{Caprini:2018mtu} (gray shaded region). The recent NANOGrav observation~\cite{NANOGrav:2023gor} is also shown in the upper left corner, which can in principle be fitted in our model for a keV-scale $M_{Z'}$ with $g'\simeq 0.4$; this parameter space is however excluded by low-energy laboratory constraints~\cite{Ilten:2018crw, Bauer:2018onh}. There is a whole suite of proposed GW experiments at various frequencies (from nHz to kHz), such as SKA~\cite{Weltman:2018zrl}, GAIA/THEIA~\cite{Garcia-Bellido:2021zgu}, MAGIS~\cite{MAGIS-100:2021etm}, AION~\cite{Badurina:2019hst}, AEDGE~\cite{AEDGE:2019nxb}, $\mu$ARES~\cite{Sesana:2019vho}, LISA~\cite{LISA:2017pwj}, TianQin~\cite{TianQin:2015yph}, Taiji~\cite{Ruan:2018tsw}, DECIGO~\cite{Seto:2001qf}, BBO~\cite{Corbin:2005ny}, ET~\cite{Punturo:2010zz},  CE~\cite{Reitze:2019iox}, as well as recent proposals for high-frequency GW searches in the MHz-GHz regime~\cite{Aggarwal:2020olq, Berlin:2021txa, Herman:2022fau, Bringmann:2023gba}. The projected sensitivities of a selected subset of these planned experiments are shown in Fig.~\ref{fig:GWplot} by the dashed/dot-dashed curves. The experiments above the mHz frequency are the most relevant ones for us, since they will probe the region around or above electroweak-scale $M_{Z'}$, complementary to the collider searches (see Section~\ref{sec:complementary}).

In the right panel of Fig.~$\ref{fig:GWplot}$, we show the GW spectra with $v_\Phi=1$ TeV for different values of $g'$. As $g'$ increases, $\alpha$ decreases and $\beta/H_*$ increases (cf.~Fig.~\ref{fig:GWparam}). Therefore, the peak amplitude goes down, while the peak frequency slightly shifts to higher values due to the slow increase in $\beta/H_*$. This gives an upper bound on $g'$ for a given $M_{Z'}$ when we require that the GW amplitude is within the sensitivity range of a given experiment. On the other hand, as $g'$ decreases, $(\Gamma/H^4)$ at $T_*$ eventually becomes smaller than one, which no longer allows bubble nucleation. This, in turn, gives a lower limit on $g'$ for a given $M_{Z'}$ [cf. the red shaded region in Fig.~\ref{fig:GWparam} right panels] so that the first-order phase transition can happen. We will exploit this feature in Section~\ref{sec:complementary} to show the model parameter space accessible at future GW experiments.

%%%%%%%%%%%%%%%%%%%%%%%%%%%%%%%%%%
\subsection{Complementarity with other laboratory constraints}
\label{sec:complementary}
\begin{figure}[t!]
    \centering
      \includegraphics[width=0.5\textwidth]{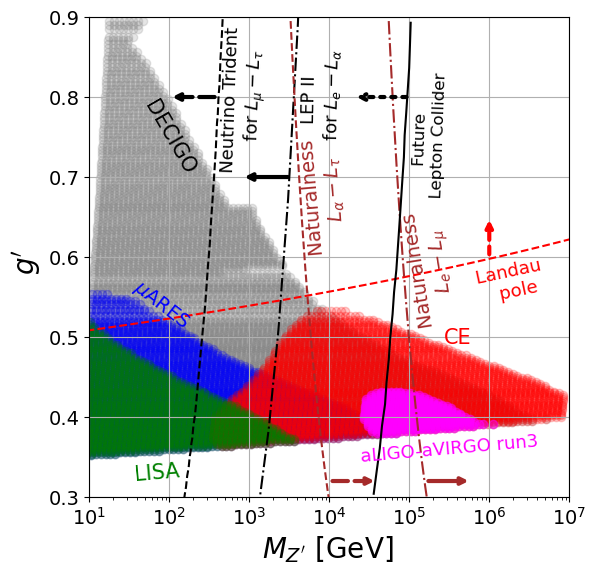}
      \caption{Discovery sensitivity (with SNR $>10$) of the future GW experiments $\mu$ARES, LISA, DECIGO and CE in the $(M_{Z'},g')$ plane. The current constraint from aLIGO-aVIRGO run 3 (with SNR $>0.1$) is also shown.  The most stringent laboratory constraints from LEP-2 and neutrino trident experiments are shown for comparison (with the arrow pointing to the exclusion region), along with the future collider sensitivity (dot-dashed line). The naturalness and perturbativity (Landau pole) constraints are also shown; see Section~\ref{sec:natural}. }
    \label{fig:SNR}
\end{figure}

%%%%%%%%%%%%%%%%%%%%%%%%%%%%%
 To estimate the stochastic GW signal strength for the ongoing GW experiments and also to obtain predictions for the future ones, we calculate the signal-to-noise ratio (SNR) $\rho$ for a given experiment by using its noise curve and by integrating over the observing time ${\rm t}_{\rm obs}$ and accessible frequency range [$f_{\min},f_{\max}$]~\cite{Allen:1996vm, 
 Allen:1997ad, Maggiore:1999vm, 
 Thrane:2013oya, Caprini:2019pxz}:
\begin{equation}
\rho = \left[n_{\rm det}{\rm t}_{\rm obs}\,\int_{f_\text{min}}^{f_\text{max}}\,\dd f\,\left(\frac{\Omega_\text{GW}(f)\,h^2}{\Omega_\text{noise}(f)\,h^2}\right)^2\right]^{1/2}\, . 
\label{eq:snr}
\end{equation}
Here $n_{\rm det}$ distinguishes between experiments aiming at detecting the GW by means of an auto-correlation ($n_{\rm det}=1$) or a cross-correlation  ($n_{\rm det}=2$) measurement. For our numerical analysis, we assume $n_{\rm det}=1$ and  
take an observation period of ${\rm t}_{\rm obs}=1$ year for each experiment. To register a detection, we demand $\rho > \rho_{\rm th}$ for some chosen threshold SNR value $\rho_{\rm th}$. With this standard, we present the potential discovery sensitivity with $\rho_{\rm th}=10$ in the plane of ($M_{Z'},\ g'$) in Fig.~\ref{fig:SNR} for three proposed experiments, namely, $\mu$ARES~\cite{Sesana:2019vho} (blue), LISA~\cite{LISA:2017pwj} (green), DECIGO~\cite{Seto:2001qf} (grey) and CE~\cite{Reitze:2019iox} (red), which are chosen for illustration in order to cover a wide frequency range, and hence, a wide range of $M_{Z'}$. 

These sensitivities are equally applicable for any flavor combination of the $U(1)_{L_\alpha-L_\beta}$ models. For comparison, we also show the most stringent laboratory constraint, which comes from LEP-2 for $U(1)_{L_e-L_\alpha}$, and from neutrino trident for $U(1)_{L_\mu-L_\tau}$ [cf.~Fig.~\ref{fig:exist}], as shown by the dashed/solid lines with the arrow pointing to the exclusion region. The best sensitivity at high $Z'$ masses coming from the Bhabha channel at $e^+e^-$ and $\mu^+\mu^-$ colliders [cf.~Fig.~\ref{fig:sensi}] is shown by the dot-dashed line. We see that future GW experiments have great potential for discovery and the theory parameter coverage corresponds up to 
$M_{Z'}\approx 4,000$ TeV. It extends the energy scale probe far beyond the collider regime, as long as the gauge coupling is sizable to produce an observable GW signal. In fact, using the existing upper limit on the GW amplitude from the third run of aLIGO-aVIRGO, we can already exclude a tiny part of the high-$M_{Z'}$ parameter space between $20-200$ TeV, as shown by the magenta region in Fig.~\ref{fig:SNR}. However, this is currently possible only for a lower SNR threshold of $\rho_{\rm th}=0.1$. 

Additional correlations between the GW signal and the collider signals could stem from the scalar sector of the model, depending on the strength of the quartic coupling $\lambda'(H^\dag H)(\Phi^\dag \Phi)$ in the scalar potential~\eqref{eq:pot}. This term induces a mixing between the SM Higgs and the extra physical scalar $\phi$. This mixing in turn induces modifications in precision electroweak observables, as well as in the SM Higgs signal strengths~\cite{Robens:2015gla, Falkowski:2015iwa}. The current LHC constraints imply that the mixing angle $\sin\theta\lesssim 0.15$ for a TeV-scale scalar~\cite{Dawson:2021jcl}, and this can be significantly improved at a future lepton collider~\cite{Buttazzo:2018qqp, AlAli:2021let}. One can also directly search for the new scalar by its decay into SM fermions, gauge bosons or Higgs pairs. For a summary of the current constraints and future prospects in the scalar mass-mixing plane, see e.g.~Refs.~\cite{Dev:2017xry, Ilnicka:2018def}. In addition, in the $U(1)$ models considered here, the scalar can decay into a pair of $Z'$ bosons or into a pair of right-handed neutrino (in extended models)~\cite{Deppisch:2018eth}, if kinematically allowed. Similarly, if the scalar and $Z'$ are light enough, they can be pair-produced from the SM Higgs decay~\cite{Deppisch:2019ldi}.  

\subsection{Theoretical constraints}
\label{sec:natural}
As the original theory adheres to classical conformal principles and is established as a massless theory, the self-energy corrections to the SM Higgs boson stem from modifications to the mixing quartic coupling $\lambda'$ in Eq.~\eqref{eq:pot}. This can be computed within the effective Higgs potential as follows:
\begin{equation}
    V_{\rm eff} \supset  \frac{\lambda'}{4}\phi^2h^2 + \frac{\beta_{\rm mix}}{8}\phi^2h^2\left(\log\left[\phi^2\right] + C\right) \, , 
\end{equation}
where the logarithmic divergence and terms not dependent on $\phi$ are all encapsulated in $C$. In the $U(1)_{L_\alpha-L_\beta}$ model we have, the principal contribution to the $\beta$-function comes from the two-loop diagrams involving the $Z'$ boson and leptons~\cite{Iso:2009nw,Oda:2017kwl}:
\begin{equation}
    \beta_{\rm mix} \supset -\frac{9g'^4m^2_\ell}{8\pi^4v^2} \, , \quad {\rm where}~m_\ell={\rm max}\{m_{\ell_\alpha},m_{\ell_\beta}\} \, .
\end{equation}
By adding a counterterm, we renormalize the coupling $\lambda'$ with the renormalization condition:
\begin{equation}
    \left. \frac{\partial^2V_{\rm eff}}{\partial h^2\partial \phi^2}\right|_{h=0, \: \phi=v_\phi} = \lambda' \, ,
\end{equation}
where $\lambda'$ is the renormalized coupling. This results in the following potential:
\begin{equation}
    V_{\rm eff} \supset  \frac{\lambda'}{4}\phi^2h^2 + \frac{\beta_{\rm mix}}{8}\phi^2h^2\left(\ln\left[\frac{\phi^2}{v^2_\Phi}\right] - 3\right) \, .
\end{equation}
Substituting $\phi=v_\Phi$, we obtain the SM Higgs mass correction as 
\begin{equation}
    \delta m^2_h = -\frac{3}{4}\beta_{\rm mix}v^2_\Phi \simeq \frac{27g'^4m_\ell^2v_\Phi^2}{32\pi^4v^2} \, .
\end{equation}
If $\delta m^2_h$ is much larger than the electroweak scale, we need a fine-tuning of the tree-level Higgs mass $m_h^2=\lambda' v_\Phi^2$ to reproduce the correct electroweak VEV. Therefore, we can introduce a fine-tuning measure as $r\equiv m_h^2/\delta m_h^2$. For instance, $r=0.1$ indicates that we need to fine-tune the tree-level Higgs mass squared at the 10\% accuracy level. This is indicated in Fig.~\ref{fig:SNR} by the solid and dashed red curves for $U(1)_{L_e-L_\mu}$ and $U(1)_{L_\alpha-L_\tau}$, respectively. The region to the right of these curves are disfavored by naturalness. However, it is worth emphasizing that these naturalness constraints are subjective ($r=0.1$ is just an arbitrary choice), and should not be treated at the same level as the {\it experimental} constraints. Nonetheless, we find from Fig.~\ref{fig:SNR} that the future colliders should be able to probe a large fraction of the $Z'$ parameter space allowed by naturalness and Landau pole constraints. 

 Another theoretical constraint that typically appears for large couplings is the perturbativity constraint. The requirement that the gauge coupling remains perturbative and does not blow up all the way up to the Planck scale is shown by the red dashed curve in Fig.~\ref{fig:SNR}, where the region above it is disfavored. This is obtained from the RG running of the gauge coupling in our model, cf.~Eq.~\eqref{eq:RGE}. We note that this constraint can be relaxed if there is some new physics between the $Z'$ scale and the Planck scale that might alter the RG evolution. 

%\BD{We also note that there exist other theoretical constraints on the $Z'$ parameter space shown here from considerations of vacuum stability and perturbativity~\cite{Das:2016zue}. Again, these are subject to the choice of the new physics cut-off scale between $v_\Phi$ and $M_{\rm Pl}$, and therefore, we do not show these in Fig.~\ref{fig:SNR}.   
%}

%%%%%%%%%%%%%%%%%%%%%%%
\section{Summary and Conclusions}
\label{sec:con}
%%%%%%%%%%%%%%%%%%%%%%%%%
In this article, we studied the phenomenology of leptophilic $Z'$ gauge bosons around the electroweak-scale mass range in the anomaly-free $U(1)_{L_\alpha-L_\beta}$ models at the future high energy $e^+e^-$ and $\mu^+\mu^-$ colliders, as well as at  gravitational wave observatories. The two independent parameters of the model are the mass of the \zp boson ($M_{Z'}$) and its coupling to the leptons ($g'$). 
We first summarized the existing bounds on the model parameters from low energy to collider searches in Section~\ref{sec:bounds}. As depicted in Fig.~\ref{fig:exist}, a large parameter space with $M_{Z^\prime} \gtrsim 100$ GeV and $g'$ up to ${\cal O}(0.1)$ 
is still unexplored. 

In Section~\ref{sec:pheno}, we analyzed in great detail the $Z'$ phenomenology at future lepton colliders. We considered a representative collider center-of-mass energy of 3 TeV~\cite{CLICdp:2018cto, MuonCollider:2022xlm}. For the considered parameter space, the \zp decays promptly. We considered various production channels such as resonant production of \zp via the radiative return, production in association with an observable photon, and the scenario where \zp is produced via the radiation of the final-state lepton.
As shown in the plots of Fig.~\ref{fig:Ecm} and Fig.~\ref{fig:xsec3TeV}, the resonant production of \zp is characterized by a resonance peak at $\sqrt{s}\simeq M_{Z^\prime}$. For $\sqrt{s}>M_{Z^\prime}$ the ISR effect facilitates the on-shell \zp production. 
The SM processes that mimic the signal are dilepton production via $s$- or 
$t$-channel exchange of the photon and $Z$ boson, as well as a 4-lepton final state via electroweak VBF, where two forward-backward leptons remain  undetected. The VBF processes have a sizable contribution to the SM background, particularly in the off-resonance region. The signal shows distinct kinematic features in the invariant dilepton mass $M_{\ell\ell}$, the rapidity $y_{\ell\ell}$, and the cosine angle of the final state leptons, which are used to discriminate the $Z'$ signal from the SM background. 
We presented the $2\sigma$ sensitivity limit in $(M_{Z'},g')$ plane as shown in Fig.~\ref{fig:sensi}. We find that  couplings down to $g^\prime \sim 10^{-3}$ can be probed for $M_{Z^\prime}\lesssim\sqrt{s}$, whereas in the off-shell regime $M_{Z^\prime}>\sqrt{s}$, the sensitivity varies between $g^\prime\simeq 0.01 - 0.1$ for $M_{Z^\prime}\simeq 3-10$ TeV.

In Section~\ref{subsec:zpgama}, we analyzed a complementary production mechanism of the \zp in association with an observable photon, {\it i.e.}, $e^+ e^-, \mu^+ \mu^-  \to Z^\prime \gamma \to \ell^+ \ell^- \gamma$ by demanding photon acceptance cuts Eq.~(\ref{eq:PSC2}). Similar to the \zp production via ISR, it exhibits resonance peak and enhancement in signal rate for $\sqrt{s}>M_{Z^\prime}$. 
It is important to note that, because of the identification of a final state photon, one can use the recoil mass $M_{\rm recoil}$ of the photon to reconstruct the \zp mass peak, regardless the decay, including the invisible mode to a neutrino pair, leading to the spectacular mono-photon plus missing energy final state. 
However, the signal rate for this case is smaller due to the radiation of an extra hard photon. 
The annihilation processes mediated  via $\gamma/Z$ are the primary background, and the VBF processes are very much suppressed in this case, as shown in Fig.~\ref{fig:xsllA} and Fig.~\ref{fig:llA3TeV}. With the similar invariant mass cut, we select the signal events and calculate the $2\sigma$ sensitivity limit. As the VBF background is suppressed, we do not use rapidity cut though there is a wide peak at $\abs{y_{Z^\prime}}$. The $2\sigma$ sensitivity limit is shown by the magenta lines in Fig.~\ref{fig:sensi}.
This channel provides the best sensitivity for the $L_e-L_\mu$ model in the mass range 100 to 300 GeV, whereas, for the $L_e-L_\tau$ and $L_\mu-L_\tau$ models, the lower reconstruction efficiency of final state $\tau$ slightly reduces the sensitivity.

We have also considered the scenarios in which the \zp does not couple to the beam leptons, and therefore, cannot be produced via direct annihilation. Instead, the \zp boson can be produced as radiation off the final state leptons. Subsequent decay of \zp to lepton pair leads to a four-lepton final state that serves as the signal for this scenario. Using the invariant mass cut as defined in Eq.~(\ref{eq:mzcut}), we  presented $2\sigma$ limit shown by red and orange curves in Fig.~\ref{fig:sensi}. Although this channel shows lower sensitivity due to a lower signal rate and a relatively larger background, it is still promising for the search in such a challenging model.

We further studied the cosmological implications of 
this model at the early universe in Section~\ref{sec:GW}. Interestingly, if these $U(1)$ models are classically conformally invariant, the phase transition at the 
$U(1)$ symmetry-breaking scale tends to be strongly first-order with ultra-supercooling, leading to 
observable stochastic gravitational wave signatures.  We calculated the GW signals in a conformal 
version of the $U(1)_{L_\alpha-L_\beta}$ models under discussion, which are complementary to our collider signals. We observed in Fig.~\ref{fig:SNR} that the GW signal generated via SFOPT occurs for 
relatively large gauge coupling $g^\prime \in [0.35, 0.55]$. Depending on the frequency range of the GW detectors, we can probe $M_{Z^\prime}$ up to several thousand TeV, well beyond the reach of colliders. In fact, the recent null results from advanced LIGO-VIRGO run 3 on the stochastic GW searches have already ruled out a small portion of the high-mass range $M_{Z^\prime} \in [ 20 ~{\rm TeV}, 1 ~{\rm PeV}]$ and gauge coupling ranging from $g^\prime \in [0.37,0.44]$.  
We concluded that more parameter space will be susceptible to future GW observatories like LISA, $\mu$ARES and Cosmic Explorer. 

 In summary, we showed the complementarity between future colliders and the GW experiments in probing the $Z'$ parameter space preferred by naturalness and Landau pole constraints. 

\noindent
\acknowledgements
 This work was supported in part by the U.S.~Department of Energy under grant Nos.~DE-SC0007914 and in part by the Pitt PACC. 
The work of BD, TH, and KX was performed partly at the Aspen Center for Physics, which is supported by the National Science Foundation under grant No. PHY-1607611 and No. PHY-2210452. 
The work of KX was also supported by the National Science Foundation under grant No. PHY-2112829, No. PHY-2013791, and No. PHY-2310497.
This work used resources of high-performance computing clusters from the Pitt CRC. The work of BD is supported in part by the U.S. Department of Energy under grant No. DE-SC0017987 and by a URA VSP Fellowship. RP acknowledges financial support for this research by the Fulbright U.S. Student Program, which is sponsored by the U.S. Department of State and
The United States – India Educational Foundation (USIEF).
\appendix
\bibliographystyle{utphys}
\bibliography{ref.bib}

\end{document}